\documentstyle[epsfig,times,amssymb,amsmath,fleqn]{mn}

\newif\ifAMStwofonts

\newcommand{\shear}{\mbox{$\mathcal{S}$}}
\newcommand{\fflex}{\mbox{$\mathcal{F}$}}
\newcommand{\gflex}{\mbox{$\mathcal{G}$}}
\newcommand{\trans}{\mbox{$\mathcal{T}$}}
\newcommand{\fflext}{\mbox{$\mathcal{F_T}$}}
\newcommand{\gflext}{\mbox{$\mathcal{G_T}$}}
\newcommand{\Ab}{\mbox{\boldmath$A$}}

\newcommand{\alphab}{\mbox{\boldmath$\alpha$}}

\newcommand{\xb}{\mbox{\boldmath$x$}}

\def\simlt{\lower.5ex\hbox{$\; \buildrel < \over \sim \;$}}
\def\simgt{\lower.5ex\hbox{$\; \buildrel > \over \sim \;$}}
\def\etal{{\it et al.}}
\def\etc{{\it etc}}
\def\ie{{\it i.e.}}
\def\eg{{\it e.g.}}
\def\cf{{\it c.f.}}

\def\pap3{Polar Shapelets}

\newcommand{\real}[1] {{\mathrm Re}\left\{#1\right\}}
\newcommand{\imaginary}[1] {{\mathrm Im}\left\{#1\right\}}

\ifoldfss
  \ifCUPmtlplainloaded \else
    \NewTextAlphabet{textbfit} {cmbxti10} {}
    \NewTextAlphabet{textbfss} {cmssbx10} {}
    \NewMathAlphabet{mathbfit} {cmbxti10} {} 
    \NewMathAlphabet{mathbfss} {cmssbx10} {} 
  \fi
  \ifAMStwofonts
    \ifCUPmtlplainloaded \else
      \NewSymbolFont{upmath} {eurm10}
      \NewSymbolFont{AMSa} {msam10}
      \NewMathSymbol{\upi}     {0}{upmath}{19}
      \NewMathSymbol{\umu}     {0}{upmath}{16}
      \NewMathSymbol{\upartial}{0}{upmath}{40}
      \NewMathSymbol{\leqslant}{3}{AMSa}{36}
      \NewMathSymbol{\geqslant}{3}{AMSa}{3E}

       \let\ge=\geqslant
    \fi
  \fi
\fi 

\ifnfssone
  \newmathalphabet{\mathit}
  \addtoversion{normal}{\mathit}{cmr}{m}{it}
  \addtoversion{bold}{\mathit}{cmr}{bx}{it}
  \newmathalphabet{\mathbfit} 
  \addtoversion{normal}{\mathbfit}{cmr}{bx}{it}
  \addtoversion{bold}{\mathbfit}{cmr}{bx}{it}
  \newmathalphabet{\mathbfss} 
  \addtoversion{normal}{\mathbfss}{cmss}{bx}{n}
  \addtoversion{bold}{\mathbfss}{cmss}{bx}{n}
  \ifAMStwofonts
    \ifCUPmtlplainloaded \else
      \UseAMStwoboldmath
      \makeatletter
      \new@mathgroup\upmath@group
      \define@mathgroup\mv@normal\upmath@group{eur}{m}{n}
      \define@mathgroup\mv@bold\upmath@group{eur}{b}{n}
      \edef\UPM{\hexnumber\upmath@group}
      \new@mathgroup\amsa@group
      \define@mathgroup\mv@normal\amsa@group{msa}{m}{n}
      \define@mathgroup\mv@bold\amsa@group{msa}{m}{n}
      \edef\AMSa{\hexnumber\amsa@group}
      \makeatother
      \mathchardef\upi="0\UPM19
      \mathchardef\umu="0\UPM16
      \mathchardef\upartial="0\UPM40
      \mathchardef\leqslant="3\AMSa36
      \mathchardef\geqslant="3\AMSa3E

       \let\ge=\geqslant
    \fi
  \fi
\fi 

\ifnfsstwo
  \DeclareMathAlphabet{\mathbfit}{OT1}{cmr}{bx}{it}
  \SetMathAlphabet\mathbfit{bold}{OT1}{cmr}{bx}{it}
  \DeclareMathAlphabet{\mathbfss}{OT1}{cmss}{bx}{n}
  \SetMathAlphabet\mathbfss{bold}{OT1}{cmss}{bx}{n}
  \ifAMStwofonts
    \ifCUPmtlplainloaded \else
      \DeclareSymbolFont{UPM}{U}{eur}{m}{n}
      \SetSymbolFont{UPM}{bold}{U}{eur}{b}{n}
      \DeclareSymbolFont{AMSa}{U}{msa}{m}{n}
      \DeclareMathSymbol{\upi}{0}{UPM}{"19}
      \DeclareMathSymbol{\umu}{0}{UPM}{"16}
      \DeclareMathSymbol{\upartial}{0}{UPM}{"40}
      \DeclareMathSymbol{\leqslant}{3}{AMSa}{"36}
      \DeclareMathSymbol{\geqslant}{3}{AMSa}{"3E}

       \let\ge=\geqslant
    \fi
  \fi
\fi 

\ifCUPmtlplainloaded \else
  \ifAMStwofonts \else 
    \def\upi{\pi}
    \def\umu{\mu}
    \def\upartial{\partial}
  \fi
\fi
\def\gs{\mathrel{\raise1.16pt\hbox{$>$}\kern-7.0pt 
\lower3.06pt\hbox{{$\scriptstyle \sim$}}}}         
\def\ls{\mathrel{\raise1.16pt\hbox{$<$}\kern-7.0pt 
\lower3.06pt\hbox{{$\scriptstyle \sim$}}}}         
\newcommand{\Psh}[1]{{P^{\rm sh}_{#1}}}

\title{Weak gravitational shear and flexion with polar shapelets}
\author[Massey \etal]
{Richard~Massey$^{1}$, Barnaby~Rowe$^{2}$, Alexandre~Refregier$^{3}$,
David~J.~Bacon$^{2}$ \& Joel~Berg\'e$^{3}$
\\
$^1$ California Institute of Technology, 1200 E.\ California Blvd., Pasadena,
     CA 91125, U.S.A.; {\tt rjm@astro.caltech.edu} \\
$^2$ SUPA (Scottish Universities Physics Alliance),
     Institute for Astronomy, Blackford Hill, Edinburgh EH9 3HJ, U.K.
     \\
$^3$ Service d'Astrophysique, B\^{a}t. 709, CEA Saclay, F-91191
     Gif sur Yvette, France
      \\}

\date{Accepted ---. Received ---; in original form 28 September 2006.}

\pagerange{\pageref{firstpage}--\pageref{lastpage}}
\pubyear{2006}

\begin{document}

\maketitle

\label{firstpage}

\begin{abstract}  

We derive expressions, in terms of ``polar shapelets'', for the image distortion
operations associated with weak gravitational lensing. Shear causes galaxy
shapes to become elongated, and is sensitive to the second derivative of the
projected gravitational potential  along their line of sight; flexion bends 
galaxy shapes into arcs, and is
sensitive to the third derivative. Polar shapelets provide a natural
representation, in which both shear and flexion transformations are compact.
Through this tool, we understand progress in several weak lensing methods. We
then exploit various symmetries of shapelets to construct a range of shear
estimators with useful properties. Through an analogous investigation, we also
explore several flexion estimators. In particular, some of the estimators can be
measured simultaneously and independently for every galaxy, and will provide
unique checks for systematics in future weak lensing analyses. Using simulated
images from the Shear TEsting Programme (STEP), we show that we can recover
input shears with no significant bias. A complete software package to
parametrize astronomical images in terms of polar shapelets, and to perform a
full weak lensing analysis, is available on the world wide web.
\end{abstract}

\begin{keywords}
methods: data analysis --- techniques: image processing ---
gravitational lensing
\end{keywords}

\section{Introduction} \label{intro}

Weak gravitational lensing is a powerful method to map the
distribution of mass in the Universe, regardless of its nature or
state (for reviews see Mellier 1999; Bartelmann \& Schneider 2001;
Refregier 2003). The apparent shapes of background galaxies become
distorted as their light travels near mass concentrations along their
line of sight to the Earth. The well-known shearing of galaxies, in
which intrinsically circular sources would be seen as elongated
ellipses, is induced by an amount proportional to the second derivative
of the projected foreground gravitational potential. Such distortion has been
measured around individual galaxy clusters (\eg\ Wittman \etal\ 2001;
Wittman \etal\ 2003; Bacon \& Taylor 2003; Brada\v{c} \etal\ 2005;
Wittman \etal\ 2006) and, in a statistical fashion, by large scale
structure (recent measurements include Massey \etal\ 2004; Van
Waerbeke \etal\ 2005; Heymans \etal\ 2005; Jarvis \etal\ 2006;
Hoekstra \etal\ 2006; Semboloni \etal\ 2006; Hetterscheidt \etal\ 
2006; Schrabback \etal\ 2006; Kitching \etal\ 2007; Massey \etal\ 2007b).

A higher-order effect, known as ``flexion'', is also emerging as a
probe of the distribution of mass on small scales, and particularly in
the inner cores of galaxy clusters (Goldberg \& Natarajan 2002; Irwin
\& Schmakova 2003; Goldberg \& Bacon 2005; Bacon \etal\ 2006; Irwin \&
Schmakova 2006; Okura, Umetsu \& Futamase 2006; Goldberg \& Leonard
2007). Variation in the shear signal across the width of a background
galaxy causes bending in its apparent shape. This is the next term in
a lensing expansion that leads towards the formation of an arc, as in
strong lensing. The flexion is sensitive to the third derivative of
the projected gravitational potential.

Precise image analysis techniques are required to detect weak
gravitational lensing, because the shapes of galaxies are changed by
the effect by only a few percent. In fact, the lensing contribution to
the shape is about an order of magnitude smaller than the dispersion
of galaxies' intrinsic morphologies and the spurious distortions
introduced by typical imperfections in telescopes. The
widely used shear measurement method by Kaiser, Squires \& Broadhurst 
(KSB, 1995) has been successful in many contexts, but
contains several documented shortcomings: it is found to be
insufficiently accurate to measure shears with a desired accuracy of
less than 1\% (\cf\ Bacon \etal\ 2001; Erben \etal\ 2001; Heymans
\etal\ 2005 (STEP1); Massey \etal\ 2007a (STEP2)), and it is
mathematically ill-defined for realistic Point Spread Functions 
(\cf\ Kaiser 2000; Kuijken 1999; Hirata \& Seljak 2003).

Several new shear measurement methods are being developed, to fully
exploit future space-based weak lensing surveys with HST or the
proposed SNAP, DUNE or JDEM missions, and ground-based wide-field
surveys such as those with Megacam, CTIO DES, VISTA darkCAM,
Pan-STARRS and LSST. A review of the various shear measurement methods
is found in STEP2, along with their division into
``active'' and ``passive'' categories. Active techniques work by
modelling galaxies as intrinsically circular, then shearing the models
until they most closely match the observed ellipticities. Passive
methods work by measuring the apparent ellipticities of objects as
well as higher order shape statistics, which are used to calibrate the
ellipticities.

Flexion measurement methods are still in relative infancy. Initial
attempts to mathematically describe the flexion distortion
(Goldberg \& Natarajan 2002; Irwin \& Shmakova 2003) were formidably
complicated. A passive estimator has been constructed by Okura, Umetsu
\& Futamase (2006), and further expanded by Goldberg \& Leonard (2007). A
completely different, probabilistic approach is taken by Irwin \&
Shmakova (2006) and Irwin, Shmakova \& Anderson (2006). However,
several important features in these approaches remain to be developed, and
they remain mathematically complex; it is therefore desirable to
find a formalism which allows maximum physical insight into the
problem. An advance towards this was made by Goldberg \& Bacon (2005),
who related flexion to the formalism of Cartesian shapelets (Refregier
2002; Refregier \& Bacon 2002). Shapelets contain all the mechanics
necessary to deconvolve galaxies and flexion estimators from the
effects of a PSF. The active method of Goldberg \& Bacon (2005) and
Goldberg \& Leonard (2007) has been used to successfully detect the
flexion signal. The mathematics has a simpler form, although it is
still not as elegant as possible.

Here, we present the image manipulations of lensing theory in terms of the
``polar shapelets'' formalism (Refregier 2003; Massey \& Refregier 2005). This
suggests a complete, orthonormal set of basis functions into which any galaxy
shapes can be decomposed. It also provides a neat way to deconvolve arbitrary
galaxy shapes from an arbitrarily complicated PSF, so we can set out under the
assumption that this problem is solved. Polar shapelets then provide a natural
representation for both shear and flexion operations, with simple mathematical
forms that yield transparent physical interpretation. The complex number
approach used throughout polar shapelets matches very conveniently with the
complex ellipticity notation of Blandford (1991) now ubiquitous in shear
literature, and with the complex formalism of flexion developed by Bacon \etal\ 
(2006). A complete software package to decompose images into polar shapelets is
available from the shapelets web site\footnote{{\tt
http://www.astro.caltech.edu/$\sim$rjm/shapelets}}.

We then exploit the inherent symmetries of polar shapelets to explore a
comprehensive range of passive measurement methods for both shear and flexion. 
To create a shear or flexion estimator, we simply need to find a combination of
shapelet coefficients that has the desired properties under each
transformation.  We generally keep the estimators as close as possible to linear
in the image, to minimise both noise and bias in the final result. The shapelet
methodology resembles a continuation of the KSB method to higher order. However,
the inclusion of higher order shape information, and a complete parametrization
of galaxy morphology, provides several new opportunities to improve on KSB, and
to remove its instabilities. Some of the shear and flexion estimators that we
describe are also independent, and can be obtained simultaneously for each
galaxy. These will provide invaluable new cross-checks for systematics in the
data analysis, which are unique to this method, and can also be combined to
increase the overall ratio of signal to noise.  As we shall discuss, one of the
shear estimators has already been proved highly successful in a blind test on
simulated images containing an applied shear, as part of the STEP programme
(Heymans \etal\ 2006; Massey \etal\ 2007a). We defer detailed testing of the
remainder until the next STEP cycle.

This paper is organised as follows. In \S\ref{lensing} we describe the shapelet
decomposition and the action of weak gravitational lensing in shapelet space. In
\S\ref{shear_estimators} we derive several possible weak shear estimators, and
discuss the performance of a key estimator on the simulated STEP images. In
\S\ref{flexion_estimators} we derive several possible weak flexion estimators. 
We conclude in \S\ref{conclusions}.

\section{Weak gravitational lensing in polar shapelet space} \label{lensing}

We shall first describe the action of weak shears and weak flexions in
polar shapelet space. This is seen as a mixing of power between an
object's various shapelet coefficients, or equivalently how much those
coefficients change under each operation. To first order, a vector of
shapelet coefficients is acted upon by simple matrices that contain
small mixing components in their off-diagonal terms. For example, a
shear takes some power from the circular ($m=0$) shapelet coefficients
and redistributes it into the elliptical ($m=2$) shapelet
coefficients, turning a circle into an ellipse.

The effect of shear as an abstract coordinate transformation has
already been derived in Cartesian shapelet space by Refregier (2003),
and in polar shapelet space by Massey \& Refregier (2005). Here, we
review this shear in the physical context of weak gravitational
lensing. Operators to perform flexion have been derived in Cartesian
shapelet space by Goldberg \& Bacon (2005). Here, we translate those
results into polar shapelet space, where they become much simpler. The
flexion operators fit naturally into the complex notation of polar
shapelets. Furthermore, the two distinct types of flexion identified
by Bacon \etal\ (2006) mix distinct sets of polar shapelet
coefficients, which can be separated elegantly.

\subsection{Polar shapelet space in the absence of lensing}
\label{sec:absenceoflensing}

\begin{figure} \centering \epsfig{figure=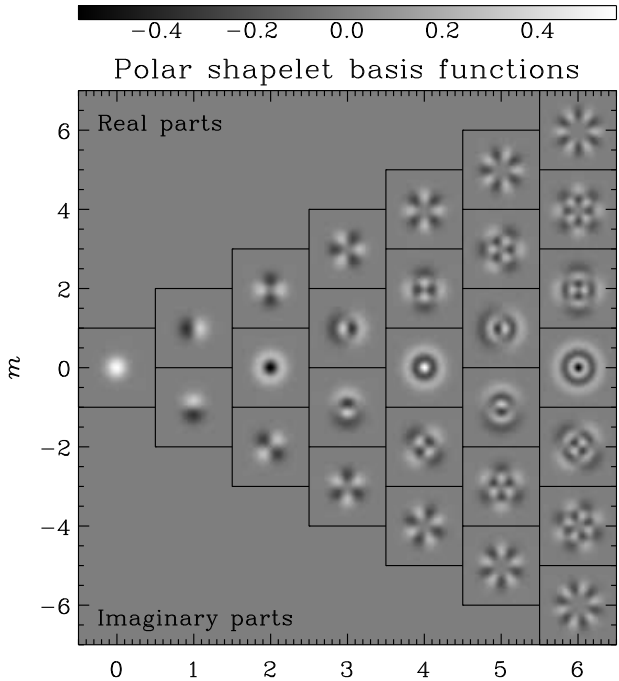,width=84mm}
\caption{The polar shapelet basis functions, with indices $n$ and $m$
that describe the number of radial and tangential oscillations. The
functions are complex, but several symmetries exist to ensure that a
reconstructed image is wholly real, and these have been used to
condense the plot. Basis functions (and shapelet coefficients) with
opposite signs of $m$ are complex conjugate pairs. Only the real part
is shown here for basis functions with $m\ge0$ and only the imaginary part
for those with $m<0$. The basis functions with $m=0$ are wholly
real. Units of the colour scale assume that $\beta=1$. 
The boxes have also been enlarged into the spaces between
allowed coefficients for clarity. \label{fig:basis_functions}}
\end{figure}

The observed image of every galaxy $f(r,\theta)$ can be decomposed
into a sum of (complex) orthogonal 2D basis functions
\begin{eqnarray}
\chi_{n,m}(r,\theta) = 
  \frac{(-1)^{\frac{n-|m|}{2}}}{\beta^{|m|+1}}
  \left[\frac{\left(\frac{n-|m|}{2}\right)!}{\pi \left(\frac{n+|m|}{2}\right)! }\right]^{\frac{1}{2}}
  ~\times~~~~~~~~~~~~~~~~~~ \\
  r^{|m|}
  L_{\frac{n-|m|}{2}}^{|m|} \left(\frac{r^2}{\beta^2}\right)
  e^\frac{-r^{2}}{2\beta^2}
  e^{-im\theta} ~. \nonumber
\end{eqnarray}
weighted by (complex) shapelet coefficients $f_{n,m}$
\begin{equation} \label{eqn:sseriesp}
f(r,\theta) = \sum_{n=0}^\infty \sum_{m=-n}^{n} f_{n,m}
\chi_{n,m}(r,\theta) ~.
\end{equation}
\noindent The basis functions, which are illustrated in
figure~\ref{fig:basis_functions}, are fully described in Massey \&
Refregier (2005) and Bernstein \& Jarvis (2002). They are Laguerre
polynomials in $r$ multiplied by sines and cosines in $\theta$, and a
circular Gaussian of width $\beta$. This scale size is chosen to match
the observed size of each galaxy, and the functions are placed at the
galaxy's centre of light. The shape of each galaxy can then be
completely described by the array of its shapelet coefficients
$f_{n,m}$. These are complex numbers, with $f_{n,-m}=f^*_{n,m}$.  The
indices $n$ and $m$ correspond to the numbers of radial and tangential
oscillations respectively: $n$ can take any nonnegative integer, and
$m$ can take any integer between $-n$ and $n$, in steps of two. The
index $m$ will be the most significant in this paper, because
coefficients with the same value of $m$ describe features of a galaxy
with the same degree of rotational symmetry.

In practice, the shapelet expansion must be truncated, and we typically use coefficients with $n$ less
than some maximum amount or, conveniently in this context, $n+|m|$ less than some amount. The latter,
``diamond''-shaped truncation scheme is a cut in the {\it total} number of oscillations, so is more
consistent with arguments concerning information content in Fourier space, like the $\theta_{\rm min}$
and $\theta_{\rm max}$ of equation~(24) in Refregier (2003). It is also better matched to the empirically
observed distribution of power in shapelet space for typical galaxies. In
figure~\ref{fig:basis_functions}, the absolute values of coefficients with $n=7,8$ or 9 and low $|m|$
(which are not shown) would typically be higher than those towards the top-right and bottom-right of
those that are shown. For galaxy shapes this truncation scheme therefore improves the data compression
ratio, or the accuracy of image recovery using a fixed number of free parameters.

In the absence of lensing, we first assume that galaxy shapes are
randomly oriented. This must be true for a sufficiently large and
widely-separated ensemble of galaxies, if there is no preferred
direction in the universe, and if galaxies are not intrinsically
aligned. The unlensed ensemble of galaxies can not contain any
angular information, so must therefore have mean shapelet coefficients
$f_{nm}$ that obey
\begin{equation}
\label{eq:f_unlensed}
\langle f_{nm} \rangle = 0,~~{\rm if}~~ m \neq 0 ~.
\end{equation}
\noindent Thus, only the $m=0$ coefficients of the ensemble average
are populated. This is the only information available about an unlensed galaxy
ensemble. It encodes the galaxies' flux
\begin{equation} \label{eqn:fdef}
F \equiv \iint f(r,\theta) ~r{\mathrm d}r{\mathrm d}\theta = \beta \sqrt{4\pi} ~ \sum_{n}^{\rm even} f_{n0}~,
\end{equation}
\noindent and radial profile (see Massey \& Refregier 2005), including their 
average size
\begin{eqnarray} \label{eqn:r2def}
R^{2} \equiv \frac{1}{F}\iint r^2f(r,\theta) ~r{\mathrm d}r{\mathrm d}\theta = \frac{ \beta^{3} \sqrt{16\pi} }{F}~ \sum_{n}^{\rm even} (n+1) ~ f_{n0}
\end{eqnarray}
\noindent and higher order shape moments like
\begin{eqnarray} \label{eqn:r4def}
\xi \equiv \iint r^4f(r,\theta) ~r{\mathrm d}r{\mathrm d}\theta = 
\beta^{5} \sqrt{64\pi} ~ \sum_{n}^{\rm even} (n^2+2n+2) ~ f_{n0} ~,
\end{eqnarray}
\noindent as defined by Okura, Umetsu\ \& Futamase (2006). All of these will be used later.

Although the following quantities will be zero on average for the population,
for each galaxy we can also define an unweighted centroid 
\begin{equation} \label{eqn:xc}
x_{c}
\equiv \frac{1}{F}\iint re^{i\theta} f(r,\theta) ~r{\mathrm d}r{\mathrm d}\theta
= \frac{ \beta^{2} \sqrt{8\pi} }{F}~ \sum_{n}^{\rm odd} \sqrt{n+1} ~ f_{n1} ~,
\end{equation}
\noindent ellipticity
\begin{eqnarray} \label{eqn:unweightede}
\varepsilon & \equiv & \frac{1}{FR^2}\iint r^2e^{2i\theta} f(r,\theta) ~r{\mathrm d}r{\mathrm d}\theta \nonumber \\
 & = & \frac{ \beta^{3} \sqrt{16\pi} }{FR^2} \sum_{n}^{\rm even} \sqrt{n(n+2)} ~f_{n2}
\end{eqnarray}
\noindent and trefoil
\begin{eqnarray} \label{eqn:trefoil}
\delta & \equiv & \frac{1}{\xi}\iint r^3e^{3i\theta} f(r,\theta) ~r{\mathrm d}r{\mathrm d}\theta \nonumber \\
 & = & \frac{\beta^4 \sqrt{32\pi}}{\xi} ~ \sum_{n}^{\rm odd} \sqrt{(n-1)(n+1)(n+3)} ~f_{n3} ~,
\end{eqnarray}
\noindent the numerator of which is the $\beta$-invariant quantity $Q$
obtained by setting $s=4$ and $m=3$ in equations~(56) and (58) of
Massey \& Refregier (2005).

\subsection{Effect of shear in shapelet space}

As a bundle of light rays from a distant galaxy passes through a
foreground gravitational field characterized by
the lensing potential $\Psi(x,y)$, the rays are
differentially deflected, and the apparent shape of the galaxy is
distorted (\cf\ Bartelmann \& Schneider 2001). The shape of the galaxy
$f(x,y)$ is sheared by an amount
\begin{equation}
\gamma\equiv\gamma_1+i\gamma_2 =
\frac{1}{2}\left(\frac{\partial^2\Psi}{\partial x^2}-\frac{\partial^2\Psi}{\partial y^2}\right)
+i\frac{\partial^2\Psi}{\partial x\partial y} ~.
\end{equation}

Positive values of the real part, $\gamma_1$, correspond to
elongations of the galaxy along the $x$-axis and compressions along
the $y$-axis. Positive values of the imaginary part, $\gamma_2$,
correspond to elongations of the galaxy along the line $y=x$ and
compressions along the line $y=-x$.  In both cases, negative values
indicate the opposite. This complex shear notation (and an analogous
form of complex ellipticity) is useful in weak lensing because both
components are expected to be zero on average in the absence of
a signal. In this case, a modulus-argument
form for shear would have a zero modulus, but no well-defined
angle. The complex form also arises very naturally in polar shapelet
space.

As shown in Massey \& Refregier (2005), under a weak lensing shear
$\widehat{\shear}$ to first order, the shapelet coefficients $f_{nm}$
transform as
\begin{eqnarray} \label{eqn:opshear}
\widehat{\shear}:f_{n,m} \rightarrow f_{n,m}^\prime = f_{n,m} ~~~~~~~~~~~~~~~~~~~~~~~~~~~~~~~~~~~~~~~~~~~~~~~~~~~~~~ \\
+ \frac{\gamma}{4}\left\{ \sqrt{(n+m)(n+m-2)}~f_{n-2,m-2} \right. ~~~~~~~~~ \nonumber \\
- \left. \sqrt{(n-m+2)(n-m+4)}~f_{n+2,m-2} \right\} \nonumber \\
+ \frac{\gamma^*}{4}\left\{ \sqrt{(n-m)(n-m-2)}~f_{n-2,m+2} \right. ~~~~~~~~~ \nonumber \\
- \left. \sqrt{(n+m+2)(n+m+4)}~f_{n+2,m+2} \right\} \nonumber
\end{eqnarray}
\noindent where the asterisk denotes complex conjugation.
For an intrinsically circular galaxy, or a galaxy ensemble whose
unlensed coefficients $\langle f_{nm} \rangle$ obey
equation~(\ref{eq:f_unlensed}), the lensed coefficients $\langle
f_{nm}' \rangle$ are left unchanged
\begin{equation}
\langle f_{n,m}' \rangle \simeq \langle f_{n,m} \rangle ~~~{\rm if}~~m\neq \pm 2 ~,
\end{equation}
\noindent except for the $|m|=2$ modes, where
\begin{equation}
\label{eq:fprime_lens}
\langle f_{n,2}' \rangle
\simeq \frac{\sqrt{n(n+2)}}{4}
\left\langle f_{n-2,0} - f_{n+2,0} \right\rangle \gamma ~,
\end{equation}
\noindent with $n=2,4,6, \ldots$. After lensing, the galaxy has
nonzero $m=0$ and $|m|=2$ coefficients (but no
others). Figure~\ref{fig:mixing} illustrates the action of mixing
between nearby shapelet coefficients. The most obvious consequence is
that the galaxy's unweighted ellipticity~(\ref{eqn:unweightede})
also becomes non-zero. However, the fractional amount by
which it changes depends upon the galaxy's radial profile. This idea
will be explored in \S\ref{shear_estimators}, along with other
combinations of combinations of $m=2$ coefficients.

Note that even a pure shear to first order can change the size of a galaxy, if it is not
intrinsically circular. But propagating series~(\ref{eqn:r2def}) through
operation~(\ref{eqn:opshear}), and comparing the result to
series~(\ref{eqn:unweightede}), it is easy to deduce that
\begin{equation} \label{eqn:changeinr2}
\widehat{\shear}:R^2 \rightarrow R^{2\prime}
 = R^2 (1 + \gamma\varepsilon^* + \gamma^*\varepsilon)
 = R^2 (1 + 2\gamma_1\varepsilon_1 + 2\gamma_2\varepsilon_2) ~.
\end{equation}
\noindent In fact, there are (only) two different linear combinations of
shapelet coefficients that are invariant under a first order shear:
\begin{eqnarray}
\Gamma_1&=&(4\pi)^\frac{1}{2}\beta~
\sum\big(f_{0,0}+f_{4,0}+f_{8,0}+\ldots\big)
\label{shapelets:eq:shearinvariant1} \\
\Gamma_2&=&(4\pi)^\frac{1}{2}\beta~
\sum\big(f_{2,0}+f_{6,0}+f_{10,0}+\ldots\big) ~.
\label{eqn:shearinvariantquantities}
\end{eqnarray}
\noindent Furthermore, their sum is the total flux $F$, whose measurement is
also independent of the choice of scale size $\beta$.

%

\subsection{Effect of flexion in shapelet space}

If the shear field varies significantly across the width of an object,
one side is distorted more than the other, and it becomes bent into an
arclet. This effect has been dubbed ``flexion''. Building upon the
work of Goldberg \& Bacon (2005), we shall now describe the
distortions that arise from such gradients in the shear field,
$\frac{\partial\gamma}{\partial x}$. The calculations will remain in
the weak lensing regime, in the sense that no terms of order
$\gamma^2$ will be considered. However, flexion is most apparent along
lines of sight close to foreground mass concentrations, where the
shear is also likely to be strong. The more rapid falloff of a flexion
signal as a function of distance from foreground mass can be used
to probe smaller physical scales than a weak shear analysis, which 
produces relatively non-local mass reconstructions.  Bacon \etal\ 
(2006) demonstrate that it can be used to more precisely measure
substructure of dark matter halos, and their inner profile or
concentration.

\begin{figure} 
\centering \psfig{figure=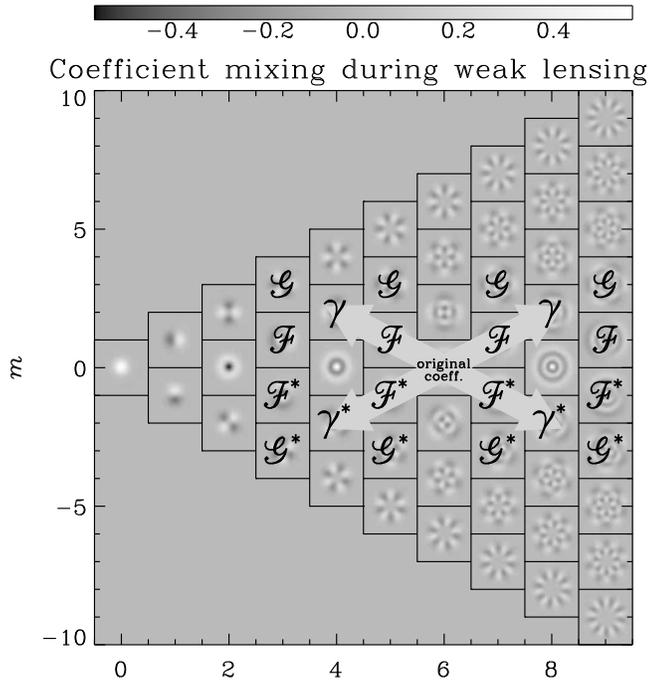,width=84mm}
\caption{ The mixing of polar shapelet coefficients under weak lensing
transformations. If a galaxy initially contains power in its $f_{6,0}$
coefficient, it will contain additional power in $f_{4,\pm2}$ and
$f_{8,\pm2}$ after shear. After both types of flexion, it will contain
additional power in eight shapelet coefficients, as illustrated. The
directions in which power moves between adjacent coefficients are the
same for a given operator wherever there are non-zero coefficients
across shapelet space, although the amount of mixing varies. Wherever
the pattern would seem to couple coefficients that do not exist, the
amount of mixing is zero.
\label{fig:mixing}}
\end{figure}

Bacon \etal\ (2006) pointed out that the flexion signal can be split into two
separate (complex) terms, the first and second flexions
\begin{eqnarray}
\fflex\equiv\left(\frac{\partial}{\partial x}-i\frac{\partial}{\partial
y}\right)\gamma =
(\gamma_{1,1}+\gamma_{2,2})+i(\gamma_{2,1}-\gamma_{1,2}) ~~ \\
\gflex\equiv\left(\frac{\partial}{\partial x}+i\frac{\partial}{\partial
y}\right)\gamma =
(\gamma_{1,1}-\gamma_{2,2})+i(\gamma_{2,1}+\gamma_{1,2}) ~.
\end{eqnarray}
\noindent We assume that these have the same units as $1/\beta$ which,
in the public code, is always expressed in terms of image pixels. Via
a derivation analogous to that in Cartesian space by Goldberg \& Bacon
(2005), we can determine the action of the flexion operators
$\widehat{\fflex}$ and $\widehat{\gflex}$ in polar shapelet
space. These are much simpler than corresponding expressions in
Cartesian shapelet space, because distinct sets of coefficients are
coupled in polar shapelet space by the two operations, and the flexion
also fits naturally into our current complex notation.
\begin{eqnarray}
\label{eqn:opfflex}
\widehat{\fflex} : f_{n,m} \rightarrow f_{n,m}^\prime = f_{n,m} ~~~~~~~~~~~~~~~~~~~~~~~~~~~~~~~~~~~~~~~~~~~~~~~~~~~~~~~~~~~ \\
~+~\frac{\fflex\beta}{16\sqrt{2}} 
   \left\{ ~ 3\sqrt{(n-m)(n+m)(n+m-2)}~f_{n-3,m-1} \right. ~~ \nonumber \\
 +~          (3n-m+10)\sqrt{(n+m)}~f_{n-1,m-1} ~~ \nonumber \\
 -~          (3n+m-4)\sqrt{(n-m+2)}~f_{n+1,m-1} ~~ \nonumber \\
 -~ \left.   3\sqrt{(n+m+2)(n-m+2)(n-m+4)}~f_{n+3,m-1}  \right\} \nonumber \\
~+~\frac{\fflex^*\beta}{16\sqrt{2}} 
   \left\{ ~ 3\sqrt{(n+m)(n-m)(n-m-2)}~f_{n-3,m+1} \right. ~~ \nonumber \\
 +~          (3n+m+10)\sqrt{(n-m)}~f_{n-1,m+1} ~~ \nonumber \\
 -~          (3n-m-4)\sqrt{(n+m+2)}~f_{n+1,m+1} ~~ \nonumber \\
 -~ \left.   3\sqrt{(n-m+2)(n+m+2)(n+m+4)}~f_{n+3,m+1}  \right\} ~, \nonumber
\end{eqnarray}
\noindent where the asterisk denotes complex conjugation. Similarly,
\begin{eqnarray}
\label{eqn:opgflex}
\widehat{\gflex} : f_{n,m} \rightarrow f_{n,m}^\prime = f_{n,m} ~~~~~~~~~~~~~~~~~~~~~~~~~~~~~~~~~~~~~~~~~~~~~~~~~~~~~~~~~~~ \\
~+~\frac{\gflex\beta}{16\sqrt{2}} 
   \left\{ ~ \sqrt{(n+m)(n+m-2)(n+m-4)}~f_{n-3,m-3} \right. ~~ \nonumber \\
 +~          \sqrt{(n+m)(n+m-2)(n-m+2)}~f_{n-1,m-3} ~~ \nonumber \\
 -~          \sqrt{(n+m)(n-m+2)(n-m+4)}~f_{n+1,m-3} ~~ \nonumber \\
 -~ \left.   \sqrt{(n-m+2)(n-m+4)(n-m+6)}~f_{n+3,m-3}  \right\} \nonumber \\
~+~\frac{\gflex^*\beta}{16\sqrt{2}} 
   \left\{ ~ \sqrt{(n-m)(n-m-2)(n-m-4)}~f_{n-3,m+3} \right. ~~ \nonumber \\
 +~          \sqrt{(n-m)(n-m-2)(n+m+2)}~f_{n-1,m+3} ~~ \nonumber \\
 -~          \sqrt{(n-m)(n+m+2)(n+m+4)}~f_{n+1,m+3} ~~ \nonumber \\
 -~ \left.   \sqrt{(n+m+2)(n+m+4)(n+m+6)}~f_{n+3,m+3}  \right\} ~. \nonumber
\end{eqnarray}
\noindent These operators are illustrated graphically in
figure~(\ref{fig:mixing}).

One crucial difference from the shear operator is that applying a
flexion shifts the galaxy's observed centroid~(\ref{eqn:xc})
by an amount
\begin{equation}
\label{eqn:flexionshift}
\Delta=\frac{R^2}{4 \beta}\left(6\fflex +
5\fflex^*\varepsilon+\gflex \varepsilon^*\right) ~,
\end{equation}
\noindent in units of $\beta$, with the real part corresponding to the
$x$ direction and the imaginary part to the $y$ direction. The
elements of expression~(\ref{eqn:flexionshift}) are easily understood
in terms of shapelet coefficients. A galaxy's centroid is constructed
from its $m=1$ coefficients. These coefficients are altered during a
first flexion $\widehat{\fflex}$ if the galaxy has power in any $m=0$
or $|m|=2$ coefficients. The $m=0$ coefficients are never all zero, so
the centroid will always shift. 
The centroid is altered during a second flexion 
$\widehat{\gflex}$ if the galaxy has power in any $|m|=2$ coefficients, 
but the effects of its $|m|=4$ coefficients happen to cancel out in
summation~(\ref{eqn:xc}). Therefore an object's ellipticity uniquely 
determines this centroid shift. No comparable
shift was introduced during shearing, so dealing with this will
present a new technical challenge for weak lensing measurement.

One mapping that will be required later is
\begin{equation} \label{eqn:changeinxi}
\widehat{\gflex}:\xi \rightarrow \xi^{\prime}
 = \xi + \gflex\rho^* + \gflex^*\rho~,
\end{equation}
\noindent where
\begin{equation} \label{eqn:rhodefinition}
\rho \equiv \beta^6\sqrt{32\pi} \sum(n+1)\sqrt{(n-1)(n+1)(n+3)}f_{n,3}~.
\end{equation}

Operators~(\ref{eqn:opfflex}) and (\ref{eqn:opgflex}) are useful for
applying an artificial flexion to an unlensed galaxy (for example,
during the manufacture of simulated images). 
However, for a practical, passive flexion measurement method, the natural 
location for the centre of a shapelet decompostion
is the post-lensing (observed) centre of light $x_c$, it
being impossible to predict the pre-lensing sky position of the source. 
This point will be
crucial in our later analysis because, for example, determinations of
ellipticity and particularly flexion depend upon the origin
of the coordinate system. To ensure that we account for this centroid
shift, we are greatly aided by the linear dependence of 
operator~(\ref{eqn:flexionshift}) upon the coefficients that will make up
our flexion estimators. The change in
coordinate frame can be simultaneously corrected for by simply
incorporating an appropriate translation in the operator used for
flexion estimation
\begin{eqnarray} \label{eqn:pracflex}
\widehat{\fflext}&\equiv&\widehat{\fflex}-\widehat{\trans}
\left(\frac{R^2}{4\beta}(6 \fflex + 5\fflex^*\varepsilon)\right)~ , \\
\widehat{\gflext}&\equiv&\widehat{\gflex}-\widehat{\trans} \nonumber
\left(\frac{R^2}{4\beta}\gflex\varepsilon^*\right) ~,
\end{eqnarray}
\noindent where, from Massey \& Refregier (2005), the translation operator is
\begin{eqnarray} \label{eqn:optranslate}
\hat{T}(\Delta):f_{n,m} \rightarrow f_{n,m}' = f_{n,m} ~~~~~~~~~~~~~~~~~~~~~~~~~~~~~~~~~~~~~~~~~~~~~~~~~~ \\
+ \frac{\Delta}{2\sqrt{2}}\left\{ \sqrt{(n+m)}~f_{n-1,m-1} \right. ~~~~~ \nonumber \\
- \left. \sqrt{(n-m+2)}~f_{n+1,m-1} \right\} ~~ \nonumber \\
+ \frac{\Delta^*}{2\sqrt{2}}\left\{ \sqrt{(n-m)}~f_{n-1,m+1} \right. ~~~~~ \nonumber \\
- \left. \sqrt{(n+m+2)}~f_{n+1,m+1} \right\} ~. \nonumber
\end{eqnarray}
\noindent These practical flexion operations for analysis of observed images
effectively isolate the observable, shape-changing part of the flexion
transformation by subtracting off the centroid shift.

As described in Goldberg \& Bacon (2005), for the purposes of
constructing workable flexion estimators the ellipticity $\varepsilon$
can be estimated from the lensed galaxy image even though 
it will itself have changed during the lensing. The change in the centroid shift this represents is small, which
can be seen from equation~(\ref{eqn:flexionshift}), and such changes will cancel on average due to the
differing rotational symmetries of $\gamma$, $\fflex$ and $\gflex$.
If deemed necessary, an estimate of the
ellipticity corrected for locally measured shear could even be used,
as there is nothing to prevent the galaxy shear analysis from being
independently performed prior to any flexion analysis.  These
operators will be used to form flexion estimators from observed galaxy
shapes in \S\ref{flexion_estimators}.

\subsection{Effect of convergence in polar shapelet space}

Convergence changes a galaxy's size and brightness. Actually measuring
convergence is difficult because galaxies are intrinsically of very
different sizes and magnitudes, and it is very hard to know what these
quantities would have been before lensing, even
statistically. (Measurements of shear and flexion are made possible by
the statistical assumption that an unlensed population of galaxies
would be round.) However, it is important to take account of the
effect of convergence on these measurements, which is given by
\begin{equation}
\kappa=\frac{1}{2}\left(\frac{\partial^2\Psi}{\partial x^2}+\frac{\partial^2\Psi}{\partial y^2}\right)
\end{equation}

Increases in apparent galaxy size potentially cause ellipticities to be
measured in different parts of a galaxy's profile -- further towards the
core or out in the wings. This is compensated for by the adaptative
choice of the shapelet scale size $\beta$ during the shapelet
decomposition described in Massey \& Refregier (2005). Indeed, the
operators $\hat{K}$ and $\hat{S}$ are commutative. Changes in galaxy
flux, or the averaging of shear estimators from bright and faint
galaxies, can be controlled by constructing estimators that are
invariant to object flux. This is trivially implemented for all of the
estimators discussed in this paper by dividing by the flux. 
To first order in $\gamma$, this quantity is invariant under a shear. It is also the most easily
measured, zeroth-order aspect of morphology: very important since this
appears on the denominator of shear estimators, where noise can translate 
into biases overall.


Note that this does not mean that the issues of ``reduced shear'' (Bartelmann \&
Schneider 2001) or indeed ``reduced flexion'' (\cf\  Okura, Umetsu \& Futamase 2006)
have been solved. Pure  gravitational shear or flexion are not observable  in
isolation. It is only possible to measure a degenerate  combination of the shear or
flexion with additional terms including the convergence.  For the unweighted shear
estimator $\hat{\gamma}_{\rm unweighted}$, which is described in
\S\ref{sec:unweighted}, the observable quantity is $\gamma/(1-\kappa)$. However, as
shown in appendix~\ref{reduced_shear}, this represents a limiting case that no longer
holds for arbitrary weighting schemes. For convenience, the observable shear
distortion will be labelled $\gamma$ hereafter in this paper; it should be understood
that this really refers not to the gravitational shear  but to the reduced shear $g$
corresponding to the estimator in question. In practice these reduced shears will be
close to the $g=\gamma/(1-\kappa)$  for the limiting unweighted case, but in  
appendix~\ref{reduced_shear} we discuss how shapelets might be used to calculate the
generalized reduced shear for each shear estimation method.


\subsection{Effect of convolution in polar shapelet space}

Galaxy shapes also change during convolution with a telescope's point spread
function. In shapelet space, convolution is another simple matrix operation
(Refregier \& Bacon 2003). Deconvolution can be performed via a matrix
inversion or simultaneously with shapelet decomposition via a method
presented in (Massey \& Refregier 2005). We shall not further discuss the
challenge of deconvolution in this paper, leaving it as a separable, and
essentially solved, problem. The main effect of deconvolution is to
correlate shapelet coefficients (since the basis functions no longer remain
completely orthogonal after convolution). The full covariance matrix can
easily be obtained during decomposition. It could, in principle, be used to
perfect the weights on coefficients in the shear estimators, although we
have derived results only in the limit where the covariance is nearly
diagonal -- which is approached by basis functions with oscillations larger
than the PSF size.


\section{Shear estimators}\label{shear_estimators}

To measure weak shear, we would like to construct some combination of
each galaxy's observed shape components that is related to the shear
field it has experienced. The combination can be of arbitrary
complexity. For individual galaxies, the measured quantity will
inevitably be noisy, because galaxies have their own intrinsic shapes,
which are changed only very slightly by weak lensing. However, we
shall aim to construct a shear estimator $\tilde{\gamma}$ for which
\begin{equation} \label{eqn:unbiased}
\langle\tilde{\gamma}\rangle = 0
\end{equation}
\noindent when averaged over a large galaxy ensemble in the absence of
shear; and, more importantly,
\begin{equation} \label{eqn:calibration}
\widehat{S}:\tilde{\gamma} \rightarrow \tilde{\gamma} + \gamma
\end{equation}
\noindent individually. As discussed in \S\ref{sec:absenceoflensing},
the first condition is easy to achieve by making sure that (the
numerator of) $\tilde{\gamma}$ contains only shapelet coefficients with
$m\ne0$. The second, calibration of the shear estimator, ensures that
the estimator is always unbiased
\begin{equation} \label{eqn:calibration2}
\langle\tilde{\gamma}\rangle = \gamma ~,
\end{equation}
\noindent but this is notoriously difficult to satisfy (\cf\ Bacon
\etal, 2001; Erben \etal, 2001; Heymans \etal, 2005; Massey \etal,
2007a). Our effort will primarily be directed here.

The easiest methodical approach towards a passive shear estimator is to
first construct a ``polarisation'' estimator $\tilde{p}$ with the same
rotational symmetries as shear. We then need to calculate its ``shear
susceptibility''
\begin{equation} \label{eqn:pgammadefinition}
P^\gamma_{ij} = \frac{\partial p_i}{\partial\gamma_j} ~,
\end{equation}
\noindent so that
\begin{equation} \label{eqn:shearpolarisation}
\widehat{S}:\tilde{p_i} \rightarrow \tilde{p_i} + P^\gamma_{ij} \gamma_j
~~\big[ + {\mathcal O}(\gamma^2) \big]~.
\end{equation}
\noindent The shear susceptibility can usefully be thought of as two
complex numbers; one for each component of shear. However, it is more
commonly expressed as a real, $2\times2$ tensor and, for the sake of
familiarity, we shall adopt that notation here. Its diagonal (real)
terms describe the amount by which the polarisation will change under a
shear. The off-diagonal (imaginary) terms describe a peculiar mixing by
which a shear in one direction can affect the polarisation in a
direction at $45^\circ$. This is introduced by complex galaxy
morphologies when a galaxy's isophotes are not concentric.

We can then construct a shear estimator
\begin{equation} \label{eqn:pshovere}
\tilde{\gamma_i} = (P^\gamma)^{-1}_{ij} \tilde{p_j} ~.
\end{equation}
\noindent to make sure that indeed
\begin{eqnarray}
\langle\tilde{\gamma_i}\rangle & = & \langle (P^\gamma_{ij})^{-1} \tilde{p_j} + (P^\gamma_{ij})^{-1}P^\gamma_{ij}\gamma_i\rangle \\
& = & \langle (P^\gamma_{ij})^{-1} \tilde{p_j} \rangle + \langle\gamma_i\rangle \\
& = & \gamma_i ~,
\end{eqnarray}
\noindent where the random intrinsic ellipticities of galaxies ensure
that the first term vanishes, and thus
condition~(\ref{eqn:calibration2}) is satisfied.

However, we immediately encounter four difficulties with shear
susceptibilities that account for most of the problems in the current
generation of shear measurement methods:
\begin{itemize}
\item {\bf ${\mathbf P}^{\mathbf \gamma}$ is noisy.} It is usually
constructed from an object's higher order shape moments, which are
even harder to measure than the polarisation. Since this appears on the
denominator, it dramatically increases the scatter of the shear
estimator: any ratio of quantities with Gaussian errors produces
the extended wings of a Cauchy distribution (as seen for a KSB analysis in
figure~2 of Massey \etal\ 2004), whose moments like $\sigma_\gamma$ do not even converge.
\item {\bf ${\mathbf P}^{\mathbf \gamma}$ is a tensor.} The matrix
inversion in equation~(\ref{eqn:pshovere}) is unstable, except for
circularly symmetric galaxies, or an unlensed population ensemble, in
which case the off-diagonal elements are always zero. In all other
cases, shearing in one direction mixes ellipticity from all other
directions, and this must be unmixed.
\item {\bf ${\mathbf P}^{\mathbf \gamma}$ is required pre-shear.} Each
galaxy is observable only after it has been lensed. Unfortunately, the
shear susceptibility factor may change during shear, to first
order in $\gamma$ for most galaxies, and to second order for even
circularly-symmetric ones.
\item {\bf The ${\mathbf P}^{\mathbf \gamma}$ formalism ignores terms of second order in
shear.} This omission may bias shear measurements at the sub-percent level
of precision,
and introduce non-linearities that depend upon an object's intrinsic
ellipticity and $|m|=4$ shapelet coefficients. 
\end{itemize}


A frequently adopted solution to the first three difficulties is to average
$P^\gamma$ from a set of intrinsically similar galaxies, or to fit a value
from a large galaxy ensemble as a function of other observables. This
approach ought to find a suitable, statistical value for all galaxies. It
diagonalises the shear susceptibility; reduces noise; and, if the population
is so large that it contains effectively no coherent shear signal, satisfies
the requirement for the pre-shear measurement. The fourth difficulty is
particularly troublesome because an object's measured ellipticity is
degenerate with the shear -- but may also be resolvable in averages over a
large population of galaxies chosen without shear-dependent biases.
Unfortunately, averaging over any large population of galaxies is inelegant,
in the sense that shear estimators for individual galaxies are no longer
self-contained. It also introduces new problems: the main issue being the
practical identification of a set of intrinsically similar galaxies. Most
observable properties of a galaxy do change during a shear, and grouping
galaxies by these leads to ``Kaiser flow'' (Kaiser, 2000). The common
challenge facing all modern shear measurement methods is to either
understand Kaiser flow statistically, or to control shear susceptibility and
thus avoid it. In appendix~\ref{kaiser_flow}, we show how measurements with
one polarisation estimator can be averaged to avoid Kaiser flow, and
maximise the weak lensing signal.

For the rest of this section, we shall construct progressively more
elaborate polarisation estimators that ameliorate the four difficulties. We
begin with simple polarisations that are compactly represented in polar
shapelet space. These still suffer from all four difficulties. We then
gradually exploit the symmetries of shapelets to add more complex features.
The process is helped by the convenient shapelet notation, although the
expressions do become more complicated. Which of these advanced shear
estimators is most appropriate to a given data set will depend on the
desired application, the image quality (for example, whether it was taken
from the ground or in space), and the number of shapelet coefficients
available for each galaxy.


%
%
%
%
%
%

\subsection{Gaussian-weighted quadrupole moment}
\label{sec:ksb}

We shall start with the simplest possible combination of shapelet
coefficients that can be used to build a polarisation estimator. Recall
that the first shapelet coefficients to be affected by a shear are
those with $|m|=2$. Like shear, these rotate as $e^{-2 \phi}$, and they
are therefore suitable for our purposes. The simplest possible
polarisation estimator is simply the first shapelet coefficient with
$m=2$, \ie\ $\tilde{p}=f_{2,2}$. This has shear susceptibility
\begin{eqnarray}
P^{\gamma}_{11} & = & (f_{0,0} - f_{4,0})/\sqrt{2} - \sqrt{3}\real{f_{4,4}} \\ 
P^{\gamma}_{22} & = & (f_{0,0} - f_{4,0})/\sqrt{2} + \sqrt{3}\real{f_{4,4}} \\
P^{\gamma}_{12} & = & P^{\gamma}_{21} ~ = ~ -\sqrt{3}\imaginary{f_{4,4}} ~.
\end{eqnarray}
%
%
%
%
\noindent In images from the Hubble Space Telescope COSMOS survey 
(Scoville \etal, 2007), for
example, $\left\langle |f_{4,4}|/f_{0,0}\right\rangle\approx0.079$,
which is not entirely negligible at the desired level of precision. By
averaging the components of $P^\gamma$ from a sufficiently large
population of observed galaxies, or fitting them as a function of other
observables like galaxy size and magnitude, we can explicitly force the
mean $m=4$ coefficients to be zero, and ensure that the measured $m=0$
coefficients are statistically corrected before shear. With this
simplification, the shear susceptibility factor can then be trivially
inverted, and we arrive at the shear estimator
\begin{equation}
\tilde{\gamma}_{\mathrm Gaussian} =
\frac{\sqrt{2}~f_{2,2}'}{\langle f_{0,0} - f_{4,0} \rangle} ~.
\end{equation}

This recovers the methods of RRG (excluding the smear correction) and
Refregier \& Bacon (2003), casting them into the more succinct framework of
polar shapelets. It recovers the $P^{sh}$ component of KSB up to the
normalisation of the polarisation estimator. To avoid biases and instability
at low signal to noise, we have chosen to keep the polarisation and shear
susceptibility linear in the image brightness. As a result however, both
quantities vary widely in the full galaxy ensemble which typically encompass
large ranges of flux and sizes, increasing the rate of Kaiser flow. A
similar decision, \ie\ whether to normalise by flux or not, will also have
to be made for all of the following shear estimators.

\subsection{Order-by-order shapelet shear estimator}

A successful shapelet decomposition contains {\it all} of the available
information about a galaxy's shape, and more information can be extracted
than that available with previous shear estimators. Since {\it all} of the
$|m|=2$ shapelet basis functions have the same rotational symmetries, each
of the corresponding shapelet coefficients can be used to form independent
(except for the covariance between shapelet coefficients after
deconvolution) polarisation estimators $p=f_{n,2}$. These have shear
susceptibilities
\begin{eqnarray}
(P^\gamma_n)_{11} & = & \frac{1}{4} \left\{
 \sqrt{n(n+2)} \left( f_{n-2,0}-f_{n+2,0} \right) \right. \\ 
 & & ~~~~~ + \sqrt{(n-4)(n-2)} \; \real{f_{n-2,4}} ~~~~~ \nonumber \\
 & & ~~~~~ - \left. \sqrt{(n+4)(n+6)} \; \real{f_{n+2,4}} \right\} \nonumber 
\end{eqnarray}
\begin{eqnarray}
(P^\gamma_n)_{22} & = & \frac{1}{4} \left\{ 
 \sqrt{n(n+2)} \left( f_{n-2,0}-f_{n+2,0} \right) \right. \\
 & & ~~~~~ - \sqrt{(n-4)(n-2)} \; \real{f_{n-2,4}} ~~~~~ \nonumber \\
 & & ~~~~~ + \left.\sqrt{(n+4)(n+6)} \; \real{f_{n+2,4}} \right\} \nonumber
\end{eqnarray}
\begin{eqnarray}
(P^\gamma_n)_{12} & = & (P^\gamma_n)_{21} \nonumber \\
  & = & \frac{1}{4} \left\{ \sqrt{(n-4)(n-2)} \; \imaginary{f_{n-2,4}} \right. ~~~~~ \\
  & & ~~~~~ - \left. \sqrt{(n+4)(n+6)} \; \imaginary{f_{n+2,4}} \right\} \nonumber ~,
\end{eqnarray}
\noindent which reduce to
\begin{equation}
P^\gamma_n=\frac{\sqrt{n(n+2)}}{4}
\langle f_{n-2,0} - f_{n+2,0} \rangle
\end{equation}
\noindent when averaged over an ensemble of galaxies as before. Thus,
for each even order $n$ available in a shapelet decomposition, we can
construct one independent, unbiased shear estimator
\begin{equation}
\tilde{\gamma}_{n} = \frac{4}{\sqrt{n(n+2)}}
\frac{f_{n,2}'}{\langle f_{n-2,0} - f_{n+2,0} \rangle} ~, 
~~{\mathrm for}~n=2,4,6,\ldots
\end{equation}
\noindent As before, these estimators are by construction unbiased,
when averaged over the galaxy population.

One way to use these additional estimators is to diagnose problems in the
measurement. Because we obtain multiple shear measurements for each galaxy during
a {\it single} PSF deconvolution, their agreement provides a strong new test of
systematics. If a pure shear signal is being successfully measured, all of the
estimators from a given galaxy should average to the same value. However, if
residual PSF effects are polluting the signal, the separate estimators will
disagree. A weak lensing pipeline must be highly robust to pass such stringent
tests, and they will provide a unique discriminatory power in future analyses.


Alternatively, the separate estimators can be linearly combined, with
arbitrary weightings
\begin{equation} \label{eqn:weighting}
p=\sum_{n=2}^{\infty} w_n f_{n,2} ~,
\label{eqn:weightedpolarisation}
\end{equation}
\noindent where the summation runs only over even indices $n$, for
only those coefficients exist. In this case
\begin{equation}
P^\gamma=\sum_{n=0}^{\infty} w_n P^{\gamma}_{n} ~.
\label{eqn:weightedsusceptibility}
\end{equation}
\noindent The weights $w_n$ can be carefully constructed to optimise
the signal-to-noise of the shear measurement (such as inverse variance
weighting, as suggested by Refregier \& Bacon 2003) or to remove
systematic biases plaguing the particular data set. For the rest of this section, we shall explore
various options for this weight function. By staying linear
in shapelet coefficients during this process, the polarisation and
susceptibility also stay linear in the image, thus preserving a
Gaussian-like distribution of estimators. In real space, changing the weights
$w_n$ is equivalent to changing the weight function used for the
polarisation estimator.


\subsection{Using galaxies' radial profiles to reduce $\sigma_\gamma$}
\label{sec:radialprof}

A galaxy's observed $m=2$ coefficients consist of intrinsic
ellipticity, shear-induced ellipticity, and noise. For an individual
galaxy, there is no way to tell what fraction of each is intrinsic,
and what fraction is the signal.  However, a shapelet decomposition contains
a great deal more information about a galaxy's morphology
that has not yet been tapped. In particular, it is the galaxy's
intrinsic radial profile ($m=0$ coefficients) that contribute most to
any change in observed ellipticity during a shear. 
Since the $m=0$ coefficients are typically much larger than any others, 
they are also, fractionally, the least changed themselves
under a small shear. We shall therefore approximate the unlensed
radial profile as the observed, measured radial profile. We can then
work out the ``radial profile'' of $m=2$ coefficients that could
possibly have been induced by lensing. Any component of the intrinsic
ellipticity that does not have the appropriate ``radial profile''
cannot possibly have been induced by lensing and, for our purposes,
can be ignored. Thus we reduce the contamination of intrinsic galaxy
ellipticity in our shear estimators, to only include components of
intrinsic ellipticity that happen to have the right profile.


We determine the required weights $w_n$ by applying a unit shear to the
rotationally-invariant part of a model, and find
\begin{equation}
\tilde{\gamma}_{\mathrm profile} \equiv
4\frac{\sum\sqrt{n(n+2)} ~ (f_{n-2,0}-f_{n+2,0}) ~ f_{n,2}}
{\big\langle \sum n(n+2)~(f_{n-2,0}-f_{n+2,0})^2 \big\rangle} ~,
\end{equation}
\noindent where one factor in the denominator comes from the shear
susceptibility factor and one from the weighted average. Of course, we
have not taken measures to eliminate the $|m|=4$ and off-diagonal
terms in the shear susceptibility factor. The shear susceptibility
will therefore need to be fitted from a galaxy population as a
function of size, magnitude and possibly radial profile. Several
shapelet-based parameters to span morphology variation are suggested
in \S7 of Masssey \& Refregier (2005).

\subsection{Diagonal shear susceptibility}
\label{sec:unweighted}

One of the difficulties with general shear estimators, as described at
the start of \S\ref{shear_estimators}, is that they require the
inversion of a (noisy) shear susceptibility
tensor~(\ref{eqn:weightedsusceptibility}). This inversion is often
unstable, and various implementations have chosen to either ignore the
off-diagonal elements, or average over a large population of galaxies
so that they disappear. The problem could be solved more easily if the
shear susceptibility were explicitly a simple scalar (times the
identity matrix) for each galaxy. Indeed, it is possible to weight the
various orders of $\tilde{\gamma}_n$ in such a way that the
off-diagonal terms in their combined susceptibility tensor from
successive orders cancel each other. The off-diagonal terms, and the
differences between the on-diagonal terms, involve $|m|=4$
coefficients that are introduced by the $\gamma^*$ terms in
equation~(\ref{eqn:opshear}). With these removed, the shear
susceptibility~(\ref{eqn:weightedsusceptibility}) will be diagonal and only involve
terms with $m=0$. This can be trivially inverted.


A simple calculation to obtain the desired $w_n$ yields
\begin{equation}
p = 4\sqrt{\pi} \beta^{3} \sum_{n}^{\infty} \sqrt{n(n+2)}~f_{n2} ~,
\end{equation}
\noindent where the prefactor has been added to reproduce familiar
quantities. In fact, $p=FR^2\varepsilon$, a version of the (radially)
unweighted ellipticity without size (or flux) normalisation. This can not
normally be calculated from images because background noise makes 
the real-space integrals diverge. A shapelet decomposition removes noise
by acting as a prior on the permitted physical properties of a
galaxy shape. This polarisation has shear susceptibility
\begin{equation}
P^\gamma=
16\pi^{\frac{1}{2}} \beta^{3} \sum_{n}^{\infty} \sqrt{n+1}~f_{n0} = 2FR^2 ~,
\end{equation}
\noindent where the right hand side refers to quantities measured {\em before}
shearing. The susceptibility is the size and magnitude of galaxies, in a curious
contrast to the previous shear susceptibilities that needed to be ensemble averaged
and fitted as a function of those observables. Furthermore, as shown in 
equation~(\ref{eqn:changeinr2}), to first order in $\gamma$, the size $R^2$ changes
under a shear in a way that  affects the overall shear estimator
\begin{equation} \label{eqn:unweightedshearest}
\widehat{\shear}:\frac{p}{2FR^2} \rightarrow \frac{p^\prime}{2F^\prime R^{2\prime}}
 = \frac{p + 2FR^2\gamma}{2FR^2(1 + \gamma\varepsilon^* + \gamma^*\varepsilon)} ~.
\end{equation}
\noindent Ensemble averaging, and expanding to first order in $\gamma$, we recover
\begin{equation}
\left\langle\frac{p^\prime}{2F^\prime R^{2\prime}}\right\rangle = 
\left\langle\frac{p}{2FR^2}\right\rangle + \gamma\Big(1-\frac{\langle\varepsilon^2\rangle}{2}\Big)
\end{equation}
\noindent with the same ``shear responsivity'' factor of 
$1-\frac{\langle\varepsilon^2\rangle}{2}$ that appears in
equation~(\ref{eqn:oneminusesquared}). Thus we obtain an unbiased shear estimator
\begin{equation} \label{eqn:changeinpoverfrsquared}
\tilde{\gamma}_{\mathrm unweighted} \equiv
\frac{\sum \sqrt{n(n+2)}f_{n2}}
{\Big(2-\langle\varepsilon^2\rangle\Big) ~ \sum\sqrt{(n+1)} f_{n0}} 
\end{equation}
\noindent that is written in terms of observable quantities alone, and requires 
minimal averaging of shapelet coefficients from a population of galaxies.

\begin{figure} 
\centering \psfig{figure=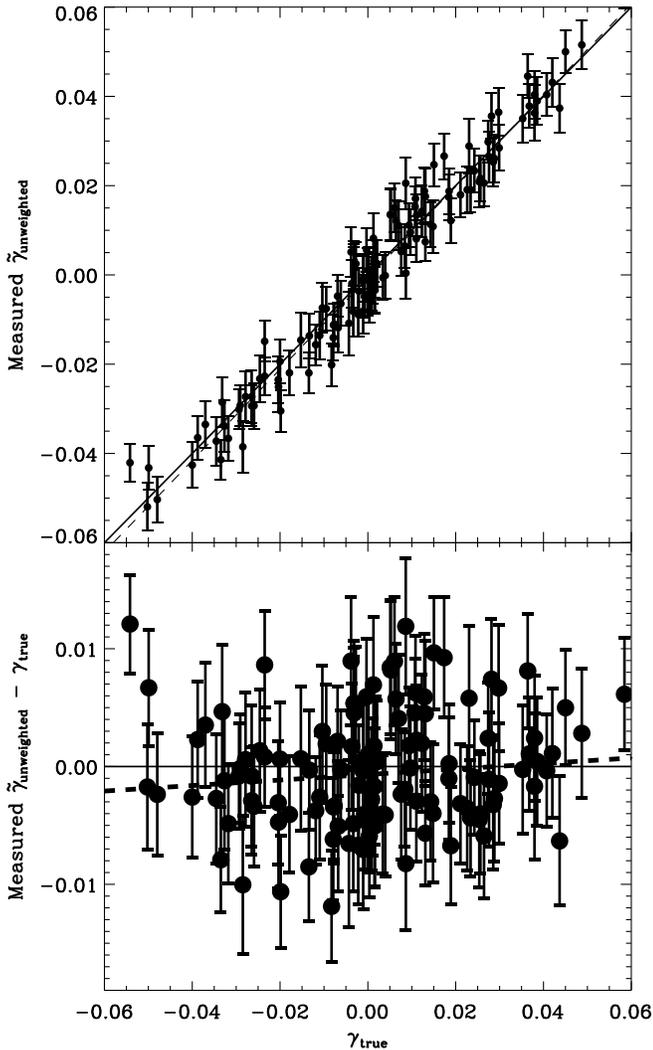,width=84mm}
\caption{Performance of the shear estimator with an explicitly diagonal
shear susceptibility tensor, on simulated images containing a known
true shear. Points show both components of the mean shear, measured in
$7\arcmin\times7\arcmin$ patches of sky where the input shear was
held constant. For a perfect shear estimator, these would lie on the
solid line. The dashed line is a linear fit to deviations from it.
\label{fig:shearinout}}
\end{figure}

This particular shear estimator emerged as one of the most successful shear
measurement methods during blind tests as part of the STEP programme (Massey
\etal\ 2007a). This programme constructed simulated images that exhibit all the
statistical properties of real astronomical images, but contain a known shear
signal. While the measurement was performed (by JB), these input shears were
kept hidden. They were then revealed publicly after all the pipelines had been
run. Figure~\ref{fig:shearinout} shows the impressive performance of our
$\tilde{\gamma}_{\mathrm unweighted}$ shear estimator for STEP2 image set A, which
was specifically designed to mimic deep {\it Suprime-Cam} images from the {\it
Subaru} telescope (Miyazaki \etal, 2002). A linear fit to these results shows a
shear calibration factor (multiplicative measurement bias) of $m=0.023\pm0.029$
for the real component of shear, and $m=0.053\pm0.029$ for the imaginary
component. There is no significant residual shear offset (additive measurement
bias), with the fitted values being $c=(-6.8\pm6.5)\times10^{-4}$ for the real
component of shear and $c=(1.3\pm6.6)\times10^{-4}$ for the imaginary component.
This estimator demonstrated the best performance throughout the STEP2 project
and, as suggested by that analysis, we shall reduce our error bars before the
next round by incorporating a galaxy weighting scheme into our weak lensing
pipeline.

One additional benefit of this estimator is that both the polarisation
and the shear susceptibility are independent of the shapelet scale
scale $\beta$. Although the ensemble average of any shear estimators
$\langle \tilde{\gamma}_{n} \rangle$ should always be independent of
$\beta$, in general, individual estimators $\tilde{\gamma}_{n}$ may
not be. But in the current case, once a shapelet series has converged,
$F$, $R^2$ and $\varepsilon$ combine coefficients in such as way as to not
depend upon the choice of $\beta$ (Massey \& Refregier, 2005). This
result is non-trivial: in our image decomposition pipeline, we choose
$\beta$ to optimise the image reconstruction and stabilise the PSF
deconvolution. However, this is only one possible goal. In {\sc imcat}
implementations of the KSB method, the equivalent of $\beta$ is instead
chosen to maximise the signal-to-noise for detection of the object. In
{\sc SExtractor} implementations of KSB, a scaling of {\sc SExtractor}
parameters is used. In all cases, the choice of $\beta$ will also be
affected by any applied shear that changes a galaxy's apparent size.
Whether this change is negligible depends on the specific
implementation of the algorithm to determine $\beta$ and must be tested
experimentally. In this case, it is reassuring to note that this
effect is formally absent, modulo the convergence of the series.

Since the STEP tests, we have also derived the full ``Kaiser flow''
behaviour of this estimator. This has a particularly interesting form,
and is discussed fully in appendix~\ref{kaiser_flow}.

\subsection{Shear-invariant shear susceptibility}
\label{shinv}

The shear susceptibility can be further simplified, by continuing to
add terms to the polarisation estimator. So far, we have used the
$|m|=2$ coefficients, but no others. In fact, all shapelet basis
functions with $|m|=\{2,6,10,14,\ldots\}$ contain the rotational
symmetries of shear; the higher order functions just contain
additional symmetries as well.  They can not be used as shear
estimators by themselves but, because of this, it is possible to add
them to the $|m|=2$ coefficients. The resulting polarisation estimator
will stay linear in shapelet coefficients, linear in image flux, and
keep all of the desired symmetries. Here we shall demonstrate how
higher order shapelet coefficients can be used to ``sweep'' terms in
the shear susceptibility down to $m=0$, and construct a polarisation
\begin{equation}
p=\sum_{m=2,6,10,\ldots}^{\infty}\sum_{n=2,4,6,\ldots}^{\infty} w_{n,m} f_{n,m}
\end{equation}
\noindent with any desired susceptibility factor.

We begin with the $f_{2,2}$ polar shapelet coefficient. As shown in
\S\ref{sec:ksb}, this has a shear susceptibility involving $f_{0,0}$,
$f_{4,0}$ and $f_{4,4}$. The weight on the $f_{2,2}$ coefficient in
$p$ can be used to create any desired factor in front of the $f_{0,0}$
term in $P^\gamma$. In \S\ref{sec:unweighted}, we added coefficient
$f_{6,2}$ in such a way to cancel out the $f_{4,4}$ coefficient in the
shear susceptibility. Instead, we shall now add an amount of $f_{6,2}$
(then $f_{10,2}$, $f_{14,2}$, \etc) to shape the susceptibility's
$f_{4,0}$ (and $f_{8,0}$, $f_{12,0}$) terms in any way.  Assuming
extrapolation to infinite $n$, we can thus construct a polarisation
estimator with arbitrary $m=0$ terms in its shear susceptibility
tensor.  However, it will also contain non-zero $|m|=4$ terms and
off-diagonal elements.

Now consider the $f_{6,6}$ polar shapelet coefficient. This has a
shear susceptibility involving $f_{4,4}$, $f_{8,4}$ and $f_{8,8}$
coefficients. This can be added to the polarisation with a weight
arranged so that the three $f_{4,4}$ terms in the shear susceptibility
now cancel out. The $f_{10,6}$ coefficient can then be added so that
the four $f_{8,4}$ terms cancel, and so on. This leaves `dangling'
terms in the shear susceptibility with $|m|=8$ (and due to series
truncation, in practice, high $n$). Successive additions of
$|m|=\{10,14,18\ldots\}$ terms to the polarisation can push these
terms to higher and higher $|m|$. Since the magnitude of shapelet
coefficients for real galaxies typically fall off rapidly with $n$ and
$|m|$, the contribution of any remaining dangling terms due to series
truncation decreases during this process.

A second, interwoven combination of shapelet coefficients starting
with $f_{4,2}$, $f_{8,2}$ and $f_{8,6}$ can also be constructed, to
make a completely separate polarisation whose shear
susceptibility involves arbitrary contributions of only $f_{2,0}$,
$f_{6,0}$, $\ldots$ coefficients.

\begin{table}
\centering 
\caption{Coefficient weights for the real part of the
$\tilde{\gamma}_{\mathrm shear-invariant}$ shear estimator
(equation~\ref{eqn:shinv}). The imaginary part of this shear estimator
involves complex conjugates of terms with every other value of $m$,
but is most conveniently found by instead rotating the galaxies by
$45^\circ$ then calculating the real part a second time.}
\begin{minipage}{140mm} 
\begin{tabular}{rrl} 
\hline
\hline
$n$ & $m$ & $w_{n,m}$ \\
\hline
2   & 2 & $\sqrt{2}$   \\
4   & 2 & $\sqrt{2/3}$ \\
6   & 2 & $\sqrt{4/3}$ \\
8   & 2 & $\sqrt{4/5}$ \\
10  & 2 & $\sqrt{6/5}$ \\
12  & 2 & $\sqrt{6/7}$ \\
14  & 2 & $\sqrt{8/7}$ \\
16  & 2 & $\sqrt{8/9}$ \\
18  & 2 & $\sqrt{10/9}$ \\
\hline
\hline
\end{tabular}
\begin{tabular}{rrc} 
\hline
\hline
$n$ & $m$ & $w_{n,m}$ \\
\hline
6    & 6  & $8/(3\sqrt{20})$    \\
8    & 6  & $4/(3\sqrt{35})$    \\
10   & 6  & $8/(3\sqrt{35})$    \\
12   & 6  & $8/(3\sqrt{105})$   \\
14   & 6  & $8/(3\sqrt{42})$    \\
10   & 10 & $16/(15\sqrt{7})$   \\
12   & 10 & $16/(15\sqrt{77})$  \\
14   & 10 & $96/(15\sqrt{462})$ \\
14   & 14 & $64/(7\sqrt{858})$  \\
\hline
\hline
\end{tabular}
\end{minipage}
\label{tab:shearinvariant} 
\end{table}

We are now free to decide the most suitable form for the shear
susceptibility, and can construct any appropriate
polarisation estimator. It would be easiest to satisfy
requirement~(\ref{eqn:calibration}) if $P^\gamma$
did not change under shear. We could then use the observed
(post-shear) value for each individual galaxy. The only two
quantities~(\ref{shapelets:eq:shearinvariant1}) and
(\ref{eqn:shearinvariantquantities}) like this can clearly be
constructed from these two interwoven combinations of shapelet
coefficients. We shall construct the polarisation estimator with shear
susceptibility $\Gamma_1+\Gamma_2=F$, where $F$ is the flux. 
Clearly, any shear estimator must
eventually be normalised so that the same shape is calculated for two
(otherwise identical) galaxies of different brightnesses. The flux is
the simplest, and most robustly measured normalisation, and the
obvious choice to put on the denominator; \ie 
\begin{equation}
\tilde{\gamma}_{\mathrm shear-invariant} \equiv
\frac{1}{F}\sum_n\sum_m w_{n,m} f_{n,m} ~.
\label{eqn:shinv}
\end{equation}

Calculating $w_{n,m}$ for this case is an elementary but tedious procedure using the iterative method
described above. The first terms are listed in table~\ref{tab:shearinvariant}. Note that although each
term in the series is well-defined, a real-space calculation in Appendix~\ref{most_linear} demonstrates
that the summation approaches a solution that is not continuously differentiable.

\subsection{Convergence issues}

Unweighted ellipticities (and higher-order properties) can not
usually be measured from real images, due to the presence of noise
that makes the integrals diverge. However, this becomes possible with
shapelets (or {\sc Im2shape} or {\sc GalFit}) because a noise-free
(and unpixellated and deconvolved) model of the galaxy is
reconstructed. For different weight functions, the question becomes
one of the convergence of galaxies' radial profiles to zero at large
radii, and the suitable truncation of its measured shapelet series. A galaxy with an
exponential profile has polar shapelet coefficients $f_{n,0}\propto
n^{-2.5}$ for a well chosen $\beta$: for a sufficiently high $n_{\rm max}$, 
all of the shear estimators do converge. For a further discussion of the 
convergence of shapelet series, see Massey \& Refregier (2005).

Furthermore, although polar shapelet coefficients can be measured to
infinite $n$ using the linear overlap method (see Massey \& Refregier
2005) for idealised data, this is not the case using the now
standard $\chi^2$ fitting method (see also Kuijken 2006; Nakajima \&
Bernstein 2006), or in the presence of pixellisation or a PSF (Berry,
Hobson \& Withington 2004). We therefore need to ensure that
sufficient coefficients can be measured, particularly for the 
elaborate polarisation estimators that converge more slowly. They may
therefore require galaxies of slightly higher signal to noise and
$n_{\rm max}$. This can be achieved by raising the magnitude or size
cut in a lensing catalogue, although the extent to which this is
necessary has to be determined by experiment. However, the more
elaborate polarisation estimators have correspondingly simpler shear
susceptibilities, which converge faster. As was the case for noise,
the minimisation of truncation errors is particularly important in the
denominator, and this may in fact prove to be the deciding factor.

\subsection{Active shear estimators with polar shapelets}

Bernstein \& Jarvis (2002), Kuijken (2006) and Nakajima \& Bernstein (2006) suggest a
rather different philosophy for constructing shear estimators. Instead of measuring an
observed polarisation, and calculating how that would have changed during shear,
Bernstein \& Jarvis (2002) shear objects (or their coordinate system) until they
appear circular. Kuijken (2006) assumes that objects were intrinsically circular 
(\ie\ models with $f_{n,m}=0$ $\forall$ $m\ne0$), then shears them until they most
closely resemble the data. This makes for a simpler calculation because the shape of
each object, including its higher order moments, is known well {\it before} the
operation. Forward shearing of an image is also useful, because it can be performed to
arbitrarily high order in $\gamma$, addressing the fourth concern in
\S\ref{shear_estimators}. In real space, the sheared ellipticity of the shapelet basis
functions in Bernstein \& Jarvis (2002) can be chosen from a continuous range; in
shapelet space, operation~(\ref{eqn:opshear}) can be exponentiated to include higher
order terms. However, a shear susceptibility factor is still needed (the calibration
factor $\mathcal{R}$ in equation~(5.33) of Bernstein \& Jarvis (2002) has the same
origin as that in our equation~(\ref{eqn:unweightedshearest})). This too must be
fitted or interpolated from a galaxy population as a function of other observables, so
offers no advantage over a shear susceptibility.

At first sight, this shear measurement philosphy seems more efficient than the passive, moment-based shear
estimators considered in the rest of this paper. Active methods do not require the full model of each
deconvolved galaxy shape, but extract only the required quantity $\gamma$. However, the decadence of
obtaining a full shape reconstruction can provide extra information that is invaluable. For example,
checking that the model's residual image is consistent with noise can indicate potential problems, and
which shear estimates to trust, in a way that is not possible if ony a single number is obtained for each
galaxy. In principle, it is possible to expand the definition of ``circularity'' in Bernstein \& Jarvis
(2002) to involve different shapelet coefficients, but this does not generate extra information that is
necessarily useful for systematic cross-checks. In that case, each fit would require a separate non-linear
iteration to find the best-fitting parameters $x_c$, $\beta$, $n_{\rm max}$ and $\gamma$, and therefore
could be subject to independent biases. Since altering the shear estimator could change any systematic
influences in this method, it would be difficult to interpret any variation between the estimators. Instead
fitting a model that is simultaneously capable of capturing all the shape information also makes the PSF
deconvolution more robust and intuitive than methods that use a model representing only the best-fit
elliptical profile of a complex galaxy shape.

In our experiments with elliptical shapelet basis functions, we have confirmed that
the choice of that ellipticity is the most unstable part of the iteration,
particularly for objects at low signal to noise. We have had one idea for a different
truncation scheme with highly elliptical basis sets. A problem arises when
fluctuations along the minor axis become smaller than the PSF or a pixel and therefore
non-orthogonal. Simply decreasing $n_{\rm max}$ (Nakajima \& Bernstein 2006) also
shortens the reach of the basis funtions along the major axis, and therefore could
potentially bias a shear measurement. However, it is possible to first rotate the
basis functions so that $\theta=0$ lies along the major axis of the ellipse, then
truncate the basis functions at different values of $n_1$ and $n_2$, the Cartesian
shapelet indices in the $x-$ and $y-$directions. If the newly-truncated coefficients
are kept, but initially set to zero, operation~(37) from Massey \& Refregier (2005)
can be used to recover the coefficients that would have been obtained from an
unrotated set of elliptical basis functions and thus continue the Bernstein \& Jarvis
(2002) method.



\vspace{-3mm}

\section{Flexion estimators}\label{flexion_estimators}

\subsection{Gaussian-weighted flexion estimators}

We shall now try to develop estimators for the weak lensing flexion, making use
of our experience with the shear estimators, and drawing tight analogies. The
simplest passive simple flexion estimator can be formed from a similar approach
to that taken when constructing our KSB-like shear estimator $\tilde\gamma_{\rm
Gaussian}$. For that, we considered the Gaussian weighted quadrupole moments,
which were the first perturbed under a shear. In this case, the shapelet
coefficients primarily affected by first and second flexions, transform as
\begin{eqnarray}
\label{eqn:ffirstorder}
\widehat{\fflext}:f_{1,1}\rightarrow f_{1,1} &+& 
\frac{\fflex \beta}{8} \left\{ 6\left(1- \frac{R^2}{\beta^2} \right)
f_{0,0} + 6\frac{R^2}{\beta^2} f_{2,0} \right. \\
 &~& \left. ~~~~~~~~~~~~~~~~-~6f_{4,0} - 5 \sqrt{2} \varepsilon^* \frac{R^2}{\beta^2}f_{2,2} 
 \right\} \nonumber \\
&+&\frac{\fflex^* \beta}{8} \left\{~ -5 \varepsilon
 \frac{R^2}{\beta^2}(f_{0,0} - f_{2,0}) \right. \nonumber \\
 &~& \left. + ~\sqrt{2}\left( 1 + 6\frac{R^2}{\beta^2} \right) 
    f_{2,2,} -  3\sqrt{6}\; f_{4,2} \right\}  \nonumber ~,
\end{eqnarray}
\begin{eqnarray}
\label{eqn:gfirstorder}
\widehat{\gflext}:f_{3,3}\rightarrow f_{3,3} &+& 
\frac{\gflex \beta}{8} \bigg\{
\varepsilon^* \frac{R^2}{\beta^2} (f_{4,2} - \sqrt{3}\;f_{2,2}) ~+  \\
&~& ~~~~~~~~~\sqrt{6}\;(f_{0,0} +f_{2,0} -f_{4,0}-f_{6,0})
 \bigg\}  \nonumber \\
&+&\frac{\gflex^* \beta}{8} \left\{2 \varepsilon \frac{R^2}{\beta^2}
 f_{4,4} - 2 \sqrt{30} \; f_{6,6} \right\} \nonumber ~.
\end{eqnarray}

Therefore, the combinations
\begin{equation}
\tilde{\fflext}=\frac{4\beta}{3}
\frac{f_{1,1}}{\langle (\beta^2-R^2)f_{0,0}+R^2f_{2,0}-\beta^2f_{4,0} \rangle}
\end{equation}
\noindent and
\begin{equation}
\tilde{\gflext}=\frac{4\sqrt{6}}{3\beta}
\frac{f_{3,3}}{\langle f_{0,0}+f_{2,0}-f_{4,0}-f_{6,0} \rangle}
\end{equation}
\noindent can be used as simple flexion estimators.

Note that the $\varepsilon$s in equations~(\ref{eqn:ffirstorder}) and
(\ref{eqn:gfirstorder}) refer to the pre-lensing ellipticity, and really
do cancel out when averaging over a population of galaxies, even in
the presence of a shear field. Changes in $R^2$ due to flexion do not bias
$\langle R^2 \rangle$ to first order either, as these cancel when averaged 
over a population of galaxies.


\subsection{Order-by-order shapelet flexion estimators}

For the small and faint galaxies that make up the majority of a weak
lensing survey, it will be difficult to measure polar shapelet
coefficients beyond the $n=6$ terms needed to estimate
$\tilde{\gflex}$ as described above. However, for those galaxies whose
higher order shapes can be measured, it is possible to generalise the
flexion estimators.

The terms in curly brackets in equations~(\ref{eqn:ffirstorder}) and
(\ref{eqn:gfirstorder}) effectively describe a flexion susceptibility
factor, which we introduce by analogy with the shear susceptibility
factor~(\ref{eqn:shearpolarisation}). We shall then be able to form
flexion estimators
\begin{equation}
\tilde{\fflex}_n \equiv \big((P^{\mathcal F}_n)_{ij}\big)^{-1} f_{n,1}
\end{equation}
\noindent and
\begin{equation}
\tilde{\gflex}_n \equiv \big((P^{\mathcal G}_n)_{ij}\big)^{-1} f_{n,m} ~.
\end{equation}
\noindent The flexion susceptibility factors are real, $2\times2$
tensors, and can be calculated using equations~(\ref{eqn:opfflex}) and
(\ref{eqn:opgflex}).  The need to subtract away an estimate of the
shift in the galaxy centroid due to the flexion itself, expressed by
equations~(\ref{eqn:pracflex}) and (\ref{eqn:optranslate}),
necessarily complicates these expressions. However, this then
describes the {\it measurable} effect of flexion on galaxy images. The
first flexion susceptibility for general $m=1$ polar shapelet
coefficients is
\begin{eqnarray}
(P^{\mathcal F}_n)_{11}+i(P^{\mathcal F}_n)_{21}~=~~~~~~~~~~~~~~~~~~~~~~~~~~~~~~~~~~~~~~~~~~~~~~~~~~~~~~~& \\
\frac{\beta}{16\sqrt{2}} \Bigg\{ 
\begin{array}{r}
3\sqrt{n+1}\Big[(n-1)(f_{n-3,0}-f_{n+1,0})~+ ~~~\nonumber \\
    (n+3)(f_{n-1,0}-f_{n+3,0})\Big]
\end{array} & \nonumber  \\
 +~ 3\sqrt{(n-3)(n-1)(n+1)} \; f_{n-3,2} & \nonumber \\
 +~ (3n+11)\sqrt{n-1}       \; f_{n-1,2} & \nonumber \\
 -~ (3n-5) \sqrt{n+3}       \; f_{n+1,2} & \nonumber \\
 -~ 3\sqrt{(n+1)(n+3)(n+5)} \; f_{n+3,2} & \nonumber \\
 +~ 2\frac{R^2}{\beta^2}(6+5\varepsilon  ) \sqrt{n+1} (f_{n+1,0}-f_{n-1,0}) & \nonumber \\
 +~ 2\frac{R^2}{\beta^2}(6+5\varepsilon^*)(\sqrt{n+3}f_{n+1,2}-\sqrt{n-1}f_{n-1,2}) & \Bigg\} \nonumber
\end{eqnarray}
\vspace{-4mm}
\begin{eqnarray}
(P^{\mathcal F}_n)_{22}+i(P^{\mathcal F}_n)_{12}~=~~~~~~~~~~~~~~~~~~~~~~~~~~~~~~~~~~~~~~~~~~~~~~~~~~~~~~~& \\
\frac{\beta}{16\sqrt{2}} \Bigg\{ 
\begin{array}{r}
3\sqrt{n+1}\Big[(n-1)(f_{n-3,0}-f_{n+1,0})~+ ~~~\nonumber \\
    (n+3)(f_{n-1,0}-f_{n+3,0})\Big]
\end{array} & \nonumber  \\
 -~ 3\sqrt{(n-3)(n-1)(n+1)} \; f^*_{n-3,2} & \nonumber \\
 -~ (3n+11)\sqrt{n-1}       \; f^*_{n-1,2} & \nonumber \\
 +~ (3n-5) \sqrt{n+3}       \; f^*_{n+1,2} & \nonumber \\
 +~ 3\sqrt{(n+1)(n+3)(n+5)} \; f^*_{n+3,2} & \nonumber \\
 +~ 2\frac{R^2}{\beta^2}(6-5\varepsilon^*) \sqrt{n+1} (f_{n+1,0}-f_{n-1,0}) & \nonumber \\
 -~ 2\frac{R^2}{\beta^2}(6-5\varepsilon  )\big(\sqrt{n+3}f^*_{n+1,2}-\sqrt{n-1}f^*_{n-1,2}\big) & \Bigg\} ~. \nonumber
\end{eqnarray}
\noindent The second flexion susceptibility for $m=3$ coefficients is
\begin{eqnarray}
(P^{\mathcal G}_n)_{11}+i(P^{\mathcal G}_n)_{21}~=~~~~~~~~~~~~~~~~~~~~~~~~~~~~~~~~~~~~~~~~~~~~~~~~~~~~~~~~~~ \\
\frac{\beta}{16\sqrt{2}} \Bigg\{ 
\begin{array}{l}
\sqrt{(n-1)(n+1)(n+3)} ~\times \nonumber  \\
~~~~~~~~~~\big(f_{n-3,0} + f_{n-1,0} - f_{n+1,0} - f_{n+3,0}\big) \nonumber \\
\end{array}  \\
+~\sqrt{(n-7)(n-5)(n-3)} \; f_{n-3,6} ~~~~~~~~~~~~~~~~~~~~ \nonumber \\
+~\sqrt{(n-5)(n-3)(n+5)} \; f_{n-1,6} ~~~~~~~~~~~~~~~~~~~~ \nonumber \\
-~\sqrt{(n-3)(n+5)(n+7)} \; f_{n+1,6} ~~~~~~~~~~~~~~~~~~~~ \nonumber \\
-~\sqrt{(n+5)(n+7)(n+9)} \; f_{n+3,6} ~~~~~~~~~~~~~~~~~~~~ \nonumber \\
+~2\frac{R^2}{\beta^2}\varepsilon   \Big(\sqrt{n-1}f^*_{n+1,2}-\sqrt{n+3}f^*_{n-1,2}\Big) ~~~~~~ \nonumber \\
+~2\frac{R^2}{\beta^2}\varepsilon^* \Big(\sqrt{n+5}f^*_{n+1,4}-\sqrt{n-3}f^*_{n-1,4}\Big) ~ \Bigg\} \nonumber
\end{eqnarray}
\vspace{-4mm}
\begin{eqnarray}
(P^{\mathcal G}_n)_{22}+i(P^{\mathcal G}_n)_{12}~=~~~~~~~~~~~~~~~~~~~~~~~~~~~~~~~~~~~~~~~~~~~~~~~~~~~~~~~~~~ \\
\frac{\beta}{16\sqrt{2}} \Bigg\{ 
\begin{array}{l}
\sqrt{(n-1)(n+1)(n+3)} ~~\times \nonumber  \\
~~~~~~~~~~\big(f_{n-3,0} + f_{n-1,0} - f_{n+1,0} - f_{n+3,0}\big) \nonumber \\
\end{array}  \\
-~\sqrt{(n-7)(n-5)(n-3)} \; f^*_{n-3,6} ~~~~~~~~~~~~~~~~~~~~~ \nonumber \\
-~\sqrt{(n-5)(n-3)(n+5)} \; f^*_{n-1,6} ~~~~~~~~~~~~~~~~~~~~~ \nonumber \\
+~\sqrt{(n-3)(n+5)(n+7)} \; f^*_{n+1,6} ~~~~~~~~~~~~~~~~~~~~~ \nonumber \\
+~\sqrt{(n+5)(n+7)(n+9)} \; f^*_{n+3,6} ~~~~~~~~~~~~~~~~~~~~~ \nonumber \\
+~2\frac{R^2}{\beta^2}\varepsilon^* \Big(\sqrt{n-1}f_{n+1,2}-\sqrt{n+3}f_{n-1,2}\Big) ~~~~~ \nonumber \\
+~2\frac{R^2}{\beta^2}\varepsilon ~ \Big(\sqrt{n+5}f_{n+1,4}-\sqrt{n-3}f_{n-1,4}\Big)
~ \Bigg\} ~.\nonumber
\end{eqnarray}

\noindent In all four cases, the last six terms are complex, and the
final two emerge from the shift in an object's apparent centroid
during flexion.

We shall now consider options by which $m=1$ and $m=3$ coefficients of
different orders $n$ can be combined. We search for sophisticated
combinations that produce flexion estimators with useful properties,
analogous to those already created for shear estimators.

\subsection{Using galaxies' radial profiles to improve flexion estimators}

Exactly as was done for shear estimators in \S\ref{sec:radialprof}, it
is possible to use knowledge of a galaxy's radial profile to restrict
which component of its $|m|=1$ or $|m|=3$ polar shapelet coefficients
could have been induced by flexion. Via a parallel derivation, we
obtain flexion estimators
\begin{equation} \label{eqn:fprofile}
\tilde{\fflex}_{\mathrm profile} \equiv 
\frac{16\sqrt{2}}{3\beta}
\frac{\sum w_n f_{n,1}}
{\Big\langle\sum(w_{n+1}^2)\Big\rangle} ~,
\end{equation}
\noindent where
\begin{eqnarray} \label{eqn:fprofilew}
w_n & = & \sqrt{n+1}(n-1)(f_{n-3,0}+f_{n+1,0}) \\
    &   & +~\sqrt{n+1}(n+3)(f_{n-1,0}+f_{n+3,0}) \nonumber \\
    &   & +~\frac{4R^2}{\beta^2}\sqrt{n+1}(f_{n+1,0}+f_{n-1,0})~; \nonumber
\end{eqnarray}
\noindent and
\begin{equation} \label{eqn:gprofile}
\tilde{\gflex}_{\mathrm profile} \equiv 
\frac{16\sqrt{2}}{\beta}
\frac{\sum w_n f_{n,3}}
{\Big\langle\sum(w_{n+3}^2)\Big\rangle} ~,
\end{equation}
\noindent where
\begin{eqnarray} \label{eqn:gprofilew}
w_n = \sqrt{(n-3)(n-1)(n+1)} ~\times~~~~~~~~~~~~~~~~~~~~~~~~~~~~~~~ \\
(f_{n-3,0}+f_{n-1,0}-f_{n+1,0}-f_{n+3,0})~. \nonumber
\end{eqnarray}

These are guaranteed to converge for a typical galaxy if sufficient terms are available in its shapelet
series. The estimator for the second flexion in particular should provide a very clean measurement with
minimal noise.


\subsection{Diagonal flexion susceptibility}

It might also be hoped that successive off-diagonal terms in the
flexion susceptibility matrices could be made to cancel via a suitable
weighting scheme $w_n$, as was possible for shear in
\S\ref{sec:unweighted}. Unfortunately, due to the presence of the
centroid-shifting correction so necessary for reliable flexion
estimators, this is difficult; especially for the first flexion.

For the second flexion we can come tantalisingly close, and indeed if
we only consider the \emph{pure} $\hat{\gflex}$ transformation of
equation (\ref{eqn:opgflex}), the weighting scheme $w_n =
\sqrt{(n-1)(n+1)(n+3)}$ can be used to form a second flexion estimator
\begin{equation} \label{eqn:puregweight}
\tilde{\gflex}_{\rm diagonal} \equiv 
\frac{2\sqrt{2}}{3\beta\mathcal{R}}\frac{\sum\sqrt{(n-1)(n+1)(n+3)}~f_{n,3}}
{\sum(n^2+2n+2)~f_{n,0}} ~.
\end{equation}
\noindent This is none other than the quantity $4\delta/3$, as developed for
HOLICS by Okura, Umetsu\ \& Futamase (2006), except for the additional ``flexion
responsivity'' factor $\mathcal{R}$. This arises because the denominator
changes during flexion (see equation~\ref{eqn:changeinxi}), biasing the overall 
estimator by an amount $1-\frac{\langle\rho\delta\rangle}{2}$ in a completely analogous
fashion to the shear responsivity factor calculated in \S\ref{sec:unweighted}.
Also, the
inclusion of terms from the flexion-induced centroid shift (\ref{eqn:pracflex})
results in off-diagonal elements in $P^{\mathcal G}$ that cannot all be removed
through any combination of the $m=3$ coefficients.

In the case of the first flexion, the prospects are worse: even if we could omit
the $\trans$ part of a practical flexion operator (which, for $\fflex$ we most 
certainly can not), a $w_n$ capable of cancelling
the off-diagonal terms in the susceptibility matrix has yet to be found by the
authors. The complication arises from the mixing of power between $\Delta m,
\Delta n= \pm 1$ coefficients in the first flexion
operation~(\ref{eqn:opfflex}). Like a centroid shift, flexion causes power to
leak between adjacent shapelet coefficients (\cf\  figure~\ref{fig:mixing}).
However, whereas the centroid shift involves only the single ladder-operator
transformations $\hat{a}^{\dagger}_r$, $\hat{a}^{\dagger}_l$, $\hat{a}_r $ and
$\hat{a}_l $ (see Refregier 2003), flexion always acts via combinations of three
of these ladder operations, taking three steps but doubling back to move only
one overall. Since $\hat{a}^{\dagger}_r$ does not commute with $\hat{a}_r $, nor
$\hat{a}^{\dagger}_l$ with $\hat{a}_l $, each $\Delta m, \Delta n= \pm 1$ term
in equation~(\ref{eqn:opfflex}) is in fact a combination of five separate
contributions, each of which representing a different, independent path between
the coefficients. Worst of all, each path contributes a differing, $n$-dependant
proportion of the overall transformation. This added level of complexity for
the first flexion transformation therefore precludes any estimator of first
flexion with vanishing off-diagonal terms in the susceptibility matrix.

The $\beta$-invariant quantity obtained by setting $s=-1$ and $m=1$ in
equations~(56) and (58) of Massey \& Refregier (2005) could be used to measure
first flexion. Unfortunately, this quantity does not appear to have any other
properties that are particularly interesting in the context of weak lensing.


\subsection{Active flexion estimators with polar shapelets}

In a similar way to the active shear estimators, it is also convenient to use a
shapelet representation when distorting a circular object (or possibly an object with
both circular $m=0$ and elliptical $|m|=2$ components) by applying flexion until it
matches the observed shape. Goldberg \& Bacon (2005) suggested a prescription in
Cartesian shapelets, which has been implemented by Goldberg \& Leonard (2007), to fit
the odd shapelet coefficients by perturbing the even ones. This results in a simple,
$\chi^2$ minimisation via matrix inversion. However, their approach is not perfectly
clean. The even Cartesian shapelet coefficients mix a great deal of structure beyond
the circularly symmetric and elliptical components. These are isolated using polar
shapelets and, furthermore, so are the first and second flexion signals. By using
polar shapelets, it is possible to fit $\fflex$ and $\gflex$ independently, from the
$|m|=1$ and $|m|=3$ polar shapelet coefficients. Since the flexion signal is spread
evenly across fewer polar shapelet coefficients than Cartesian ones, but noise from an
uncorrelated sky background is equal in all shapelet coefficients, using fewer
coefficients will result in a cleaner fit, with improved signal to noise.

\section{Conclusions} \label{conclusions}

We have described the mechanics of weak gravitational lensing in terms
of ``polar shapelets'' (Refregier 2003; Massey \& Refregier
2005). This is a set of basis functions that can be used to
parametrize images, in a way that appears convenient for both
weak shear and flexion measeurement. The symmetries of polar shapelets are
well-matched to those of lensing. For example, the complex notation of
polar shapelet coefficients, where their modulus represents the amount
of power, and their phase represents their orientation, mirrors that
commonly used in the literature to define a complex ellipticity. In
addition, polar shapelets concisely encapsulate the ideas of weak
flexion that have been recently developed.

The symmetries inherent to the polar shapelet formalism provide useful
insight into the parallel symmetries of weak lensing
distortions. We have exploited this relation to construct estimators
that are able to simultaneously extract both the weak shear and flexion
signal from the observed shapes of distant galaxies. We attempt to
bypass some of the limitations of KSB and other shear measurement
methods that were reviewed in STEP2. Adopting the
classification scheme from that programme, we briefly discussed the recasting of
alternative, ``active'' shear and flexion estimators into the polar
shapelet formalism, and more comprehensively explored the options available
for ``passive'' shear and flexion estimators. 

The Gaussian-weighted shear estimator $\tilde{\gamma}_{\rm Gaussian}$ recovers old methods like KSB and
RRG, but in a compact complex notation. The unweighted shear estimator $\tilde{\gamma}_{\rm unweighted}$
takes advantage of the noise-free shapelet reconstructions to diagonalise the shear susceptibility tensor
into a scalar quantity. This particular quantity happens to be easily derivable in real space as well. The
simplification of the shear susceptibility is completed with $\tilde{\gamma}_{\rm shear-invariant}$. With
this, the shear susceptibility is simply the object's flux: a robustly measured quantity, and one that does
not change to first order during shear. The growing complexity of these shear estimators solves
progressively more of the four issues highlighted with previous-generation shear measurement methods.
However, they also converge more slowly, and require high-$n$ coefficients to be  available. The later
estimators may consequently be accessible only on galaxy images with higher signal-to-noise. The best
estimator to use (which may not even be the same for an entire population) may therefore depend on the flux
of an object, or the nature of residual problems found in any particular dataset. One final, particularly
interesting alternative option is the estimator $\tilde{\gamma}_{\rm profile}$ that can reduce the rms
shear noise due to intrinsic galaxy ellipticities, by exploting additional information about each galaxy's
radial profile.  Analogous estimators for flexion also exist for most of these options.

Interestingly, our method permits several independent shear estimators and
several flexion estimators to be obtained simultaneously for each galaxy.
Because we calculate them following a {\it single} PSF deconvolution or
non-linear iteration, their agreement (or otherwise) will provide a stringent
new test on the PSF modelling and for other residual systematic effects. Such
tests are unique to our approach, as the interpretation of analogous active
shear estimators would be hindered by the need to perform a separate PSF
deconvolution for each estimator, and possibly removing any shared defects.

We have demonstrated the performance of one of our key shear
estimators via blind tests that were part of the Shear TEsting
Programme (STEP; Massey \etal, 2007a). We shall
compare the practical performance of our remaining shear estimators in
a future round of STEP simulations. We are also implementing an option
to input a known flexion signal in the image simulation suite of
Massey \etal\ (2004). We are planning a smaller-scale, flexion version
of STEP, to calibrate the performance of emerging weak flexion estimators,
including the ones presented in this work as well as others presented
elsewhere. In the mean time, a complete IDL software package capable of
performing the shapelet image decomposition, including the weak lensing
manipulation and analysis described in this paper, is available from
the shapelets web site {\tt
http://www.astro.caltech.edu/$\sim$rjm/shapelets}.

\section*{Acknowledgments}

The authors are pleased to thank Richard Ellis, Gary
Bernstein, Sarah Bridle, Mandeep Gill, Dave Goldberg, Catherine
Heymans, John Irwin, Mike Jarvis, Reiko Nakajima, Jason Rhodes, Nick
Scoville, Marina Shmakova and Lisa Wright for their help. DB is
supported by a PPARC Advanced Fellowship.

\appendix

\section{Reduced Shear} \label{reduced_shear}

\subsection{Idealized case -- isophotal or no weighting}
For the purposes of the following discussion we here reproduce much of
the work of Schneider \& Seitz (1995). We use $\xb_s$ and
elsewhere the suffix $s$ to denote coordinates and quantities in the
galaxy source plane, and $\xb$ and no suffix to denote coordinates
and quanitities in the lensed image plane.  Let $I_s(\xb_s)$ be
the surface brightness distribution of the source and let $W(I_s)$ be
some weighting function of the surface brightness. For the case of no 
weighting, $W(I_s)=I_s$.

We define the centre of the source by
\begin{equation}
\bar{\xb}_s \equiv \frac{\int \xb_s ~ W(I_s(\xb_s)) ~{\mathrm d}x_s{\mathrm d}y_s}
                              {\int W(I_s(\xb_s)) ~{\mathrm d}x_s{\mathrm d}y_s} ~,
\end{equation}
and the quadropole matrix of the source by
\begin{equation} \label{source}
Q^{(s)}_{ij} \equiv
\frac{\int (\Delta \xb_s)_i  (\Delta \xb_s)_j W(I_s(\xb_s)) ~{\mathrm d}x_s{\mathrm d}y_s}
     {\int W(I_s(\xb_s)) ~{\mathrm d}x_s{\mathrm d}y_s}
\end{equation}
where $\Delta \xb_s = \xb_s - \bar{\xb}_s$.
To describe the shape (including orientation) of a source, we define
the complex ellipticity,
\begin{equation}
\varepsilon_{s} = \frac{ \big(Q_{11}^{(s)} - Q_{22}^{(s)}\big) + 2 i Q^{(s)}_{12} }
{Q_{11}^{(s)} + Q_{22}^{(s)} } ~.
\end{equation}

The gravitational imaging of a general source is described by the lens equation
\begin{equation}\label{lenseq}
\xb_s = \xb - \alphab(\xb) ~.
\end{equation}
Where the mass distribution is smooth on scales of galaxy images,
the imaging of the source can be approximated by the locally 
linearized lens mapping
\begin{equation}\label{eqn:dxdy}
{\mathrm d}x{\mathrm d}y = \Ab(\xb) ~ {\mathrm d}x_s{\mathrm d}y_s ~,
\end{equation}
where
\begin{equation}
\Ab(\xb) = \left(
\begin{array}{cc}
1 - \kappa - \gamma_1 & -\gamma_2 \\
-\gamma_2 & 1- \kappa + \gamma_1
\end{array} \right) ~,
\end{equation}
the Jacobian of the lens equation \eqref{lenseq}. \newline

Now, we can also define analagous second moments $Q_{ij}$ 
for the \emph{lensed image} of a source
\begin{equation} \label{image}
Q_{ij} = \frac{\int \Delta \xb_i \Delta \xb_j W(I(\xb)) ~ {\mathrm d}x{\mathrm d}y ~,}
              {\int W(I(\xb))~ {\mathrm d}x{\mathrm d}y} ~,
\end{equation}
Using \eqref{eqn:dxdy}, and the fact that image surface brightness is conserved under
gravitational light deflection so that 
\begin{equation} \label{surfb}
W(I_s(\xb_s)) = W(I(\xb)) ~.
\end{equation}
Using the linearized mapping we may write $\Delta \xb_i
= A_{ij} (\Delta \xb_s)_j $, giving the result
\begin{equation}\label{trans}
Q^{(s)}_{ij} = A_{ik}A_{jl}Q_{kl} ~,
\end{equation}
\ie\ that $Q_{ij}$ transforms as a tensor for a locally linearized mapping.
The applicability of this 
desirable result rests heavily on the condition \eqref{surfb}.
We may write the Jacobian as
\begin{equation}
\Ab(\xb) = (1- \kappa) \left(\begin{array}{cc}
1 - g_1 & -g_2 \\
 - g_2  & 1 + g_1
\end{array}\right) ~,
\end{equation}
where we have defined the \emph{reduced shear} $g \equiv \gamma/(1 -
\kappa)$.  The transformation between $\varepsilon$ and $\varepsilon_{s}$ can then
be written as
\begin{equation}
\varepsilon_{s} = \frac{ \varepsilon - 2g + g^2 \varepsilon^* }{1 + |g^2| - 2 \real{g
  \varepsilon^*} } ~.
\end{equation}
We see immediately that the transformation between the
source and image ellipticities $\varepsilon_{s}$ and $\varepsilon$ depends
solely on the combination $g$.

Incidentally, we can continue this calculation one more step and obtain,
to first order in $g$,
\begin{equation}
\varepsilon = \varepsilon_s + 2g -2\varepsilon(\varepsilon_1g_1+\varepsilon_2g_2) ~,
\end{equation}
\noindent which, averaging over a population ensemble, is
\begin{eqnarray}
\langle\varepsilon\rangle & = & \langle\varepsilon_1\rangle + 2 \big( 1 - \langle\varepsilon_1^2\rangle \big) g_1 
                                         - 2\langle\varepsilon_1\varepsilon_2\rangle g_2 \\
                         & ~ & +~ i\Big(\langle\varepsilon_2\rangle + 2 \big( 1 - \langle\varepsilon_2^2\rangle \big) g_2 
                                         + 2\langle\varepsilon_1\varepsilon_2\rangle g_1\Big) \nonumber\\
                                   & = & \big(2 - \langle\varepsilon^2\rangle\big) g \label{eqn:oneminusesquared} ~.
\end{eqnarray}

\subsection{More general case -- weighting by a function of position}

We noted above that the tensor transformation of $Q_{ij}$ relies on
the invariance under lensing transformation of the weighted surface
brightness distribution, a condition that is only satisfied for an
isophotal weighting function $W=W(I)$.  This schema carries practical
difficulties for noisy images, and in general we wish to
weight objects by multiplying their image by a fixed function 
$W(\xb)$, such that
\begin{equation} \label{weightedimage}
Q_{ij} \equiv \frac{\int \Delta \xb_i \Delta \xb_j W(\xb) I(\xb) ~ {\mathrm d}x{\mathrm d}y}
                   {\int W(\xb) I(\xb) ~ {\mathrm d}x{\mathrm d}y} \;.
\end{equation}
It is from weighted moments such as these that weak lensing shear is
generally measured, and such moments (with gaussian $W$ functions) are
directly equivalent to combinations of the $f_{n2}$ and $f_{n0}$ polar
shapelet coefficients.

However, we see instantly that the combination $I(\xb) W(\xb)$ 
\emph{is no longer invariant under the lensing transformation}:
\begin{equation}
I(\xb) W(\xb) \neq I_s(\xb_s) W_s(\xb_s)
\end{equation}
in general. This prevents us from writing the transformation between 
$Q_{ij}$ and $Q^{(s)}_{ij}$ in the simple form of \eqref{trans}. The
quadrupole moments no longer transform as tensors and we must instead write
\begin{equation} \label{transw}
Q^{(s)}_{ij} = B_{ijkl}Q_{kl} \; \;,
\end{equation}
where each element of $B_{ijkl}$ depends upon $\gamma_1$, $\gamma_2$,
$\kappa$, and the functional forms of $I$ and $W$.
Importantly, because each element of $B_{ijkl}$ will vary with each of
these quantities, we 
cannot therefore assume that the transformation between 
$\varepsilon_{s}$ and $\varepsilon$ will depend
solely on the combination $g$ for an arbitrary weighting
function.

The differences between \eqref{trans} and \eqref{transw} are generally
assumed to be small for practical weak lensing purposes. Shapelet 
space is convenient for the calculation of elements
of $B_{ijkl}$. For a given weighting function, the
transformation may be written approximately as a power series in $\gamma_1$,
$\gamma_2$, $\kappa$, and the image moments/shapelet coefficients
$f_{nm}$.  In this way, shapelets provides one method for
estimating the \emph{generalized} reduced shear for each galaxy
image, a complicated
function of $\gamma_1$, $\gamma_2$, $\kappa$, $f_{nm}$ and $W$ in each case.


\section{Kaiser flow} \label{kaiser_flow}

\subsection{Population Response} \label{population}

In \S\ref{sec:unweighted}, we obtained expressions for unweighted,
unnormalised second moments for each galaxy. We constructed an
unnormalised size measure $q_0\equiv FR^2$ and two unnormalised
polarisation components $q_1\equiv F\varepsilon_1$ and $q_2\equiv
F\varepsilon_2$, all of which have strong flux dependencies.

We must now find an estimator for the local shear on an image given
these polarisations. If we were interested in a shear estimate for a
single galaxy, we might argue that since the lensed quantities we
measure are related to unlensed quantities by
\begin{equation}
q_0^\prime=q_0+2q_\beta \gamma_\beta
\label{qshr1}
\end{equation}
\begin{equation}
q_\alpha^\prime=q_\alpha+2q_0 \gamma_\alpha ~,
\label{qshr2}
\end{equation}
\noindent with $\alpha,\beta=1,2$, we could use an estimator for the shear
\begin{equation}
\tilde{\gamma}_\alpha=\frac{\langle q^\prime_\alpha \rangle}
                           {2 \langle q_0 \rangle} ~.
\end{equation}
\noindent However, when we come to combining shear estimators from
galaxies with different flux and size properties, this approach is not
adequate. Firstly, it does not give a prescription for how to
optimally combine the estimates from galaxies with very different flux
and shape properties. Furthermore, it ignores the fact that, under
shear, some galaxies will flow out of a cell containing galaxies at a
given flux, $q_0$ and ellipticity, while other galaxies will flow in -
and these flows may not cancel; a weighting scheme should take account
of this.

In seeking to address these issues, we closely follow the approach
offered by Kaiser (2000; section 3.2), although the fact that we are
dealing with unweighted moments simplifies our analysis.

We wish to obtain an estimator for the shear that takes into account
the shear-induced flow of galaxies in the parameter space $(F, q_0,
q^2)$, where $F$ is the flux and $q^2=q_\alpha q_\alpha$ is an
invariant measure of the ellipticity amplitude of an object. We will
find it convenient to describe $q_\alpha=q \hat{q}_\alpha$, with the
unit polarisation vector given by $\hat{q}_\alpha = \{ \cos{\phi} ,
\sin{\phi} \}$, \ie\ $\phi$ gives the position angle of the galaxy.
In this case, we can describe a volume element for polarisation by $q
{\mathrm d}q {\mathrm d}\phi$, or $\frac{1}{2}{\mathrm d}(q^2) {\mathrm d}\phi$.

Let us consider the distribution of galaxies in this parameter
space. We will represent sheared distributions as primed
quantities. If the number density in this parameter space is $n$, then
we can describe the conservation of galaxies under shear by
\begin{eqnarray}
n^\prime(F^\prime,q_0^\prime,q^{\prime2},\phi^\prime)
{\mathrm d}F^\prime {\mathrm d}q_0^\prime {\mathrm d}(q^{\prime2})
{\mathrm d}\phi^\prime= ~~~~~~~~~~~~~~~~~~~~~~~~~~~~~~~~~~\\
n(F,q_0,q^2,\phi) {\mathrm d}F
{\mathrm d}q_0 {\mathrm d}(q^2) {\mathrm d}\phi ~. \nonumber
\end{eqnarray}
\noindent Note that this differs from Kaiser's (2000) analysis in not
requiring integration over distinct polarisabilities, as these
polarisabilities are themselves given by $q_0$ and $q_\alpha$ in our
case, due to using unweighted moments.

We now multiply both sides of this equation by
$W(F^\prime,q_0^\prime,q^{\prime2}) q^\prime_\alpha = W(F,q_0+\delta
q_0, q^2+\delta q^2)(q_\alpha+\delta q_\alpha)$, where $W$ is an
arbitrary function, and integrate over all variables. This will
ultimately allow us to obtain a relation between the average
polarisation, the distribution of galaxies, and the shear. Initially we
find
\begin{eqnarray}
\label{eqn:appAeq5}
\int {\mathrm d}F^\prime {\mathrm d}q_0^\prime {\mathrm d}(q^{\prime2})
{\mathrm d}\phi^\prime n^\prime W(F^\prime, q_0^\prime, q^{\prime2}) q^\prime_\alpha =
~~~~~~~~~~~~~~~~~~~~~~~~~~~~~~~~~~~~ \\ \int {\mathrm d}F {\mathrm d}q_0 {\mathrm d}(q^2) {\mathrm d}\phi n W(F+\delta F, q_0+\delta q_0,
q^2+\delta q^2) (q_\alpha+\delta q_\alpha) ~. \nonumber
\end{eqnarray}
\noindent This can be simplified by noting that, because of the 
isotropy of the unsheared population,
\begin{equation}
\int {\mathrm d}F {\mathrm d}q_0 {\mathrm d}(q^2) {\mathrm d}\phi 
n W(F, q_0, q^2) q_\alpha =0 ~.
\label{eqn:appAeq6}
\end{equation}
\noindent Making a Taylor expansion of the left hand side of
equation~(\ref{eqn:appAeq5}), we obtain
\begin{eqnarray}
\int {\mathrm d}F^\prime {\mathrm d}q_0^\prime {\mathrm d}(q^{\prime2}) {\mathrm d}\phi' n^\prime W(F^\prime, q_0^\prime, q^{\prime2}) q^\prime_\alpha =
~~~~~~~~~~~~~~~~~~~~~~~~~~~~~~~~~~ \\ 
\int {\mathrm d}F {\mathrm d}q_0 {\mathrm d}(q^2) {\mathrm d}\phi n \left(W \delta q_\alpha+
\frac{\partial W}{\partial q_0} \delta q_0 q_\alpha +
\frac{\partial W}{\partial (q^2)} \delta (q^2) q_\alpha \right) ~. \nonumber
\end{eqnarray}

If we note from equations (\ref{qshr1}) and (\ref{qshr2}) that $\delta
q_0=2q_\beta \gamma_\beta$, $\delta q_\alpha=2q_0 \gamma_\alpha$, and $\delta
(q^2) = 4 q_0 \gamma_\beta q_\beta$, we can integrate the above expression by
parts to obtain
\begin{eqnarray}
\int {\mathrm d}F^\prime {\mathrm d}q_0^\prime {\mathrm d}(q^{\prime2}) {\mathrm d}\phi^\prime n^\prime W(F^\prime, q_0^\prime, q^{\prime2}) q^\prime_\alpha =
~~~~~~~~~~~~~~~~~~~~~~~ \\ \gamma_\beta \int {\mathrm d}F {\mathrm d}q_0 {\mathrm d}(q^2) {\mathrm d}\phi W \Big(2 n q_0
\delta_{\alpha \beta} ~~~~~ \nonumber \\
- 2 q_\beta q_\alpha
\frac{\partial n}{\partial q_0} - 4 q_0 q_\beta q_\alpha
\frac{\partial n}{\partial (q^2)} \Big) ~. \nonumber
\end{eqnarray}
\noindent Since $W(F^\prime,q^\prime_0,q^{\prime2})=W(F,q_0,q^2)$ to first order
in the shear, we can omit it on both sides (we are free to do this as it is
arbitrary), to obtain a relation between the mean of the sheared galaxies'
polarisations, and the galaxy distribution function, sizes, and shapes 
\begin{eqnarray}
\int n q_\alpha {\mathrm d}F {\mathrm d}q_0 {\mathrm d}(q^{\prime2}) {\mathrm d}\phi = 
\gamma_\beta \int \Bigg\{2 n q_0
\delta_{\alpha \beta} ~~~~~~~~~~~~~~~~~~~~~~~~~ \\ - 2 q_\beta q_\alpha \frac{\partial n}{\partial q_0} -
4 q_0 q_\beta q_\alpha \frac{\partial n}{\partial (q^2)} \Bigg\} {\mathrm d}F {\mathrm d}q_0 {\mathrm d}(q^2) {\mathrm d}\phi ~. \nonumber
\end{eqnarray}

We can usefully write this in terms of an average only over position
angles of galaxies. If we move to writing expressions in terms of the
density
\begin{equation}
n(F,q_0,q^2)=\int {\mathrm d}\phi n(F,q_0,q^2,\phi) ~,
\end{equation}
\noindent and note that the average over position angles $\langle q_\beta
q_\alpha \rangle =\frac{1}{2}q^2 \delta_{\alpha \beta}$, then we can
write the average of $q_\alpha$ over position angle only (i.e. at
fixed $F, q_0, q^2$) as
\begin{equation}
\langle q_\alpha \rangle_{F,q_0,q^2}=\gamma_\alpha \left[2q_0 -
\frac{1}{n}\frac{\partial n}{\partial q_0}-\frac{2}{n}\frac{\partial
n}{\partial (q^2)}q_0 q^2  \right] ~,
\end{equation}
\noindent where $n$ is $n(F,q_0,q^2)$ rather than $n(F,q_0,q^2,\phi)$. It will
be useful to introduce the susceptibility $P$, where $\langle q_\alpha
\rangle_{F,q_0,q^2} = P(F,q_0,q^2) \gamma_\alpha$ with
\begin{equation}
P(F,q_0,q^2)=2q_0 - \frac{1}{n}\frac{\partial n}{\partial
q_0}-\frac{2}{n}\frac{\partial n}{\partial (q^2)}q_0 q^2 ~.
\end{equation}

We have therefore obtained the appropriate polarisation to use as a
function of flux, size and shape for an ensemble of galaxies. Hence we
can construct a shear estimator for galaxies in a particular small
cell in $(F, q_0, q^2)$ space
\begin{equation}
\hat{\gamma}_\alpha^{\rm cell} = \frac{1}{N P} \sum_{\mbox{\scriptsize
gals in cell}} q_\alpha ~,
\label{cellest}
\end{equation}
\noindent where $N$ is the number of galaxies in the cell in question. However,
we would like an estimator which did not require the splitting of
galaxies into cells in parameter space, and which optimally combines
the estimators from different galaxies. We discuss this problem in the
next section.

\subsection{Optimal Weighting}

Here we will discuss the optimal weighting of shear estimators for a
spatial cell-average shear. Again, we are following the work of Kaiser
(2000). 

For our parameter-space cell shear estimate given in equation
(\ref{cellest}) above, we can find the estimator variance
\begin{equation}
\left\langle (\hat{\gamma}^{\rm cell})^2 \right\rangle = \frac{2}{N^2 P^2}
\left\langle \left( \sum q_1 \right)^2 \right\rangle = \frac{1}{N^2 P^2}
\left\langle \sum q^2 \right\rangle ~,
\end{equation}
\noindent where the final equality assumes that galaxy polarisations are
essentially uncorrelated in the weak shear regime. Thus we obtain
\begin{equation}
\left\langle (\hat{\gamma}^{\rm cell})^2 \right\rangle = \frac{q^2}{N P^2} ~.
\end{equation}

Since parameter-space cells provide shear estimates which are
uncorrelated from cell to cell, the optimal weighting $W_{\rm cell}$
is proportional to $1/\langle (\hat{\gamma}^{\rm cell})^2 \rangle = N
P^2/q^2$, as then the overall estimator variance will be minimised. So
the final total shear estimate for a small spatial aread will be given
by
\begin{eqnarray}
\hat{\gamma}_\alpha^{\rm total}
 & = & \frac{\sum_{\rm cells} (N P^2/q^2)(\sum_{\mbox{\scriptsize gals in cell}} q_\alpha / N P)}{\sum_{cells} N P^2/q^2} \nonumber \\
 & = & \frac{\sum_{\rm galaxies} Q \hat{q}_\alpha}{\sum_{\rm galaxies} Q^2} ~,
\end{eqnarray}
\noindent where $Q\equiv P/q$.


\section{The most nearly linear shape estimator in real space} 
\label{most_linear}

\subsection{Simple polarisation estimators}
\label{ksb}

The simplest polarisation estimator can be constructed for a galaxy image
$I(x,y)$ as
\begin{eqnarray}
\tilde{p}_1 & \equiv & \frac{1}{F}\iint (x^2-y^2) ~ I(x,y) ~{\mathrm d}x{\mathrm d}y \\
\tilde{p}_2 & \equiv & \frac{1}{F}\iint 2xy ~ I(x,y) ~{\mathrm d}x{\mathrm d}y ~,
\end{eqnarray}
\noindent where the flux $F \equiv \iint I(x,y) ~{\mathrm d}x{\mathrm
d}y$. The diagonal components of the shear susceptibility tensor for
this estimator take the simple form $2R^2$ (evaluated without the
weight), and we recover the shear estimator $\tilde{\gamma}_{\rm
unweighted}$ from \S\ref{shinv}. The $F$ factor could also have been
incorporated directly into the shear susceptibility factor; then both
terms are formally linear, and it is only at the point of forming a
shear estimator that any ratios need to be taken.

If we ignore the correction for PSF convolution, the KSB method can also be
recast in these simple terms. This requires Gaussian-weighted quadrupole
ellipticities
\begin{eqnarray}
\tilde{p}_1 & \equiv & \frac{1}{R^2}\iint (x^2-y^2) ~ e^{-\frac{x^2+y^2}{2r_g^2}} I(x,y) ~{\mathrm d}x{\mathrm d}y \\
\tilde{p}_2 & \equiv & \frac{1}{R^2}\iint 2xy ~ e^{-\frac{x^2+y^2}{2r_g^2}} I(x,y) ~{\mathrm d}x{\mathrm d}y ~,
\end{eqnarray}
\noindent where 
\begin{equation}
R^2\equiv \iint (x^2+y^2) e^{-(x^2+y^2)/(2r_g^2)} I(x,y) ~{\mathrm d}x{\mathrm d}y
\end{equation}
\noindent and $r_g$ is the scale size of a Gaussian weight function.
This is introduced to make sure the integrals converge in a noisy
image, and to eliminate the effects of neighbouring objects.
Unfortunately, this weight function complicates the correction for the
PSF, and makes the corresponding $\Psh{}$ tensor messy (see
equations~(5-2) to (5-4) in KSB). Introducing a ratio of moments at this
early stage reduces the dynamic range of the ellipticities and the
shear susceptibility, but also exacerbates the noise. KSB also derive a
correction for the effects of PSF convolution on a galaxy's shape, with
this Gaussian and assumptions about the form of the PSF built-in.

\subsection{General linear shape estimators}

We can generalise this procedure by defining two arbitrary weight functions
$W_i(x,y)$, with $i\in\{1,2\}$, that can be used to define two linear
polarisations
\begin{equation}
\tilde{p}_i \equiv \iint W_i(x,y)~I(x,y) ~{\mathrm d}x{\mathrm d}y
\end{equation}
\noindent The coordinate system is then distorted by a shear
\begin{equation}
\left( \begin{array}{c} x\prime \\ y\prime \end{array}\right)
=
\left( \begin{array}{cc} 
1+\gamma_1 & \gamma_2 \\
\gamma_2   & 1-\gamma_1 
\end{array}\right)
\left( \begin{array}{c} x \\ y \end{array}\right)
\end{equation}
\noindent and our shape estimators for the observed image become
\begin{eqnarray}
\tilde{p}_i = \iint W_i(x,y)~\Bigg[I(x,y) ~~~~~~~~~~~~~~~~~~~~~~~~~~~~~~~~~~~~~~~~~~~~~~~~~~~~ \\
 - \gamma_1x\frac{\partial I}{\partial x} 
 + \gamma_1y\frac{\partial I}{\partial y}
 - \gamma_2x\frac{\partial I}{\partial y} 
 - \gamma_2y\frac{\partial I}{\partial x} \Bigg] ~{\mathrm d}x{\mathrm d}y ~. \nonumber
\end{eqnarray}
\noindent Integrating each term by parts, and including a boundary
condition on the rapid convergence of the image to zero at large
radii, we obtain
\begin{eqnarray}
\label{eqn:pgamma}
\tilde{p}_i ~ = ~ p^{\mathrm{int}}_i + ~ \gamma_1 \iint \left[x\frac{\partial
W_i}{\partial x} - y\frac{\partial W_i}{\partial y}\right] I(x,y)
~{\mathrm d}x{\mathrm d}y ~~ \\ 
 + ~ \gamma_2 \iint \left[x\frac{\partial W_i}{\partial y} +y\frac{\partial W_i}{\partial x}\right] I(x,y) ~{\mathrm d}x{\mathrm d}y ~. \nonumber
\end{eqnarray}

This pair of integrals, for each of the two weight functions, make up
the four coefficients in the shear susceptibility tensor. This
procedure can also be followed in polar coordinates, where we find
\begin{eqnarray}
\label{eqn:pgammapolar}
\tilde{p}_i ~ = ~ p^{\mathrm{int}}_i ~~~~~~~~~~~~~~~~~~~~~~~~~~~~~~~~~~~~~~~~~~~~~~~~~~~~~~~~~~~~~~~~~~~~~~~~~~~~~~~~~ \\
+ ~ \gamma_1 \iint \left[r\cos{2\theta}\frac{\partial
W_i}{\partial r} - \sin{2\theta}\frac{\partial W_i}{\partial \theta}\right] I(r,\theta)
~r{\mathrm d}r{\mathrm d}\theta ~~ \nonumber \\ 
 + ~ \gamma_2 \iint \left[r\sin{2\theta}\frac{\partial W_i}{\partial r}
 + \cos{2\theta}\frac{\partial W_i}{\partial \theta}\right] I(r,\theta) ~r{\mathrm d}r{\mathrm d}\theta ~. \nonumber
\end{eqnarray}

\subsection{Shear-invariant shear susceptibility}

As was the case in section~\ref{ksb}, it is always impossible to form
a completely linear shear estimator, since bright galaxies would then
yield larger shear estimators than faint ones; it will always be
necessary to normalise a shear estimator by something proportional to
the object's flux. However, we can construct one ellipticity for which
the shear susceptibility tensor is diagonal and whose diagonal
coefficients are {\it exactly equal to} that flux (this is equivalent
to including a factor of $1/F$ in an ellipticity estimator that has
$P^\gamma\equiv1$). This will solve all three problems raised at the
beginning of \S\ref{shear_estimators}, because the flux is the most
easily measured quantity of an object, the matrix inversion can be
replaced by a division, and the flux is not changed under a shear (nor
will the overall shear estimator be changed by lensing magnification).

To achieve this shear estimator with the desired integrals from 
equation~(\ref{eqn:pgamma}), we require
\begin{eqnarray}
x\frac{\partial W_1}{\partial x} ~ - ~ y\frac{\partial W_1}{\partial y} ~~ = ~~ y\frac{\partial W_2}{\partial x} ~ + ~ x\frac{\partial W_2}{\partial y} & = & 1 \\
y\frac{\partial W_1}{\partial x} ~ + ~ x\frac{\partial W_1}{\partial y} ~~ = ~~ x\frac{\partial W_2}{\partial x} ~ - ~ y\frac{\partial W_2}{\partial y} & = & 0 ~,
\end{eqnarray}
\noindent so that
\begin{eqnarray}
\frac{\partial W_1}{\partial x} ~~ = ~~ \frac{x}{x^2+y^2} ~~~~,~~~~~
\frac{\partial W_1}{\partial y} ~~ = ~~ \frac{-y}{x^2+y^2} & \\
\frac{\partial W_2}{\partial x} ~~ = ~~ \frac{y}{x^2+y^2} ~~~~,~~~~~
\frac{\partial W_2}{\partial y} ~~ = ~~ \frac{x}{x^2+y^2} & ,
\end{eqnarray}
\noindent or in polar coordinates
\begin{eqnarray}
\frac{\partial W_1}{\partial r}      ~~ = ~~ \frac{\cos{(2\theta)}}{r} ~~~~,~~~~~
\frac{\partial W_1}{\partial \theta} ~~ = & -\sin{(2\theta)} & \\
\frac{\partial W_2}{\partial r}      ~~ = ~~ \frac{\sin{(2\theta)}}{r} ~~~~,~~~~~
\frac{\partial W_2}{\partial \theta} ~~ = & \cos{(2\theta)} & .
\end{eqnarray}
\noindent These equations are inconsistent. Therefore no continuous, analytic function exists with all the
properties desired for a fully linear shear estimator. However, a series approximation that tends to these
requirements is given by the expansion in \S\ref{shinv}.

\bsp
\label{lastpage}


\begin{thebibliography}{99}
\bibitem{bs}          Bartelmann M.\ \& Schneider P., 2001, Physics Reports 340, 2910
\bibitem{bj02}        Bernstein G.\ \& Jarvis M., 2002, AJ 123, 583
\bibitem{flexion3}    Bacon D., Goldberg D., Rowe B.\ \& Taylor A., 2006, MNRAS 365, 414
\bibitem{sheartest1}  Bacon D., Refregier A., Clowe D., Ellis R., 2001, MNRAS 325, 1065 
\bibitem{bt3d901902}  Bacon D.\ \& Taylor A., 2003, MNRAS 344 1307
\bibitem{blandford}   Blandford R., Saust A., Brainerd T.\ \& Villumsen J., 1991, MNRAS 251, 600
\bibitem{bulletsw}	  Brada\v{c}, M.\ \etal, 2005, A\&A 437, 49
\bibitem{cha01}       Chang T.\ \& Refregier, A., 2004, ApJ 617, 794
\bibitem{sheartest2}  Erben T., Van Waerbeke L., Bertin E., Mellier Y., Schneider P., 2001, A\&A 366, 717
\bibitem{flexion2}    Goldberg D.\ \& Bacon D., 2005, ApJ 619, 741
\bibitem{flexion4}    Goldberg D.\ \& Leonard A., 2007, ApJ 660, 2
\bibitem{flexion1}    Goldberg D.\ \& Natarajan P., 2002, ApJ 564, 65
\bibitem{cshet}       Hetterscheidt M.\ \etal, 2006, A\&A submitted (astro-ph/0606571)
\bibitem{cshey}       Heymans C.\ \etal, 2005, MNRAS 361, 160
\bibitem{step1}       Heymans C.\ \etal, 2006, MNRAS 368, 1323
\bibitem{reglens}     Hirata C., Seljak U., 2003, MNRAS 343, 459
\bibitem{cshoe}       Hoekstra H.\ \etal, 2006, ApJ 647, 116
\bibitem{sextupole1}  Irwin J.\ \& Shmakova M., 2003 (astro-ph/0308007)
\bibitem{sextupole2}  Irwin J.\ \& Shmakova M., 2006, ApJ 645, 1
\bibitem{sextupole3}  Irwin J., Shmakova M.\ \& Anderson J., 2006, ApJ submitted (astro-ph/0607007)
\bibitem{csjar}       Jarvis M., Jain B., Bernstein G.\ \& Dolney D., 2006, ApJ 644, 71
\bibitem{ksb}         Kaiser N., Squires G.\ \& Broadhurst T., 1995, ApJ 449, 460
\bibitem{k2k}         Kaiser N., 2000, ApJ 537, 555
\bibitem{kit3d}       Kitching T., Heavens A., Taylor A., Brown M., Meisenheimer K., Wolf C., Gray M.\ \& Bacon D., MNRAS submitted (astro-ph0610284)
\bibitem{kkpsf}       Kuijken K., 1999, A\&A 352, 355
\bibitem{kkshapelets} Kuijken K., 2006, A\&A 456, 827
\bibitem{xwht}        Massey R., Bacon D., Refregier A.\ \& Ellis R., 2005, MNRAS 359, 1277
\bibitem{shape3}      Massey R.\ \& Refregier A., 2005, MNRAS 363, 197
\bibitem{shims}       Massey R., Refregier A., Conselice C. \& Bacon D., 2004, MNRAS 348, 214
\bibitem{step2}       Massey R.\ \etal, 2007a, 376, 13
\bibitem{cosmos_cs}   Massey R.\ \etal, 2007b, ApJ~in press (astro-ph/0701480)
\bibitem{subaruobs}   Miyazaki S.\ \etal\ 2002, ApJ 580, L97
\bibitem{holics}      Okura Y., Umetsu K.\ \& Futamase T. 2006, ApJ in press (astro-ph/0607288)
\bibitem{refreview}   Refregier A., 2003, ARA\&A 41, 645
\bibitem{shape1}      Refregier A., 2003, MNRAS 338, 35 
\bibitem{shape2}      Refregier A. \& Bacon D., 2003, MNRAS 338, 48 
\bibitem{ref02}       Refregier A., Rhodes J., \& Groth E., 2002, ApJ 572, L131
\bibitem{rrg}         Rhodes J., Refregier A.\ \& Groth E., 2001, ApJ 552, L85
\bibitem{cosmos}      Scoville N.\ \etal, 2007, ApJ in press (astro-ph/0612306)
\bibitem{3point2}     Schneider P.\ \& Lombardi M., 2002, A\&A submitted (astro-ph/0207454)
\bibitem{cssch}       Schrabback T.\ \etal, 2006, A\&A submitted (astro-ph/0606611)
\bibitem{cssem}       Semboloni E.\ \etal, 2006, A\&A 452, 51
\bibitem{csvan}       Van Waerbeke L., Mellier Y.\ \& Hoekstra H., 2005, A\&A 429, 75
\bibitem{witclus01}   Wittman D., Tyson, J. A., Margoniner V., Cohen J., Dell'Antonio I., 2001, ApJ 557, L89
\bibitem{witclus03}   Wittman D., Margoniner, V., Tyson J. A., Cohen J., Becker A., Dell'Antonio I., 2003, ApJ 597, 218
\bibitem{witclus06}   Wittman D.\ \etal, 2006, ApJ 643, 128


\end{thebibliography}
\end{document}